\documentclass[11pt]{article}

\usepackage[margin=1in]{geometry}
\usepackage{latexsym}
\usepackage{graphicx}
\usepackage{float}
\usepackage{multicol,multirow}
\usepackage{amsmath,amssymb,amsfonts}
\usepackage{mathrsfs}
\usepackage{amsthm}
\usepackage{rotating}
\usepackage{appendix}
\usepackage[authoryear,round]{natbib}
\bibpunct{(}{)}{;}{a}{}{,}
\usepackage[T1]{fontenc}
\usepackage{times}
\usepackage{textcomp}
\usepackage{xcolor}
\usepackage{hyperref}
\usepackage{url}

\title{From objective discovery to prediction of global ocean eco-provinces: A pathway for trustworthy learning}

\author{
Makayla McDevitt$^{1,\ast}$ \quad Maike Sonnewald$^{1,2,3}$ \quad Stephanie Dutkiewicz$^{4}$\\[4pt]
\small $^1$University of California Davis, Davis, CA, 95616, USA\\
\small $^2$University of Washington, Seattle, WA, 98195, USA\\
\small $^3$NOAA Geophysical Fluid Dynamics Laboratory, Princeton, NJ, 08540, USA\\
\small $^4$Massachusetts Institute of Technology, Cambridge, MA, 02139, USA\\[6pt]
\small $^\ast$Corresponding author: Makayla McDevitt (\href{mailto:mrmcdevitt@ucdavis.edu}{mrmcdevitt@ucdavis.edu}), ORCID: \href{https://orcid.org/0009-0004-7035-3458}{0009-0004-7035-3458}
}
\date{}

\begin{document}
\maketitle

\begin{abstract}
Marine ecosystems are increasingly impacted by climate change, necessitating tools to identify and predict spatial habitat information. To build such tools, ecological marine provinces, ``eco-provinces," ecologically meaningful regions in the global ocean can be used. We use unsupervised machine learning (ML) to identify eco-provinces with corresponding uncertainty measures based on output of a global simulation of phytoplankton functional types. Our work aims to create a proof of concept to predict eco-provinces based on satellite ocean color data. To do so, we develop a hierarchy of explainable dense ensemble networks to infer how well the eco-provinces can be detected from modeled ocean color fields. Key results include that the delineated eco-provinces are both ecologically meaningful and can be inferred with high skill. However, no straightforward relationship was found where adding more input data to the network consistently improves inference skill, and there is an intricate tradeoff between inputs and prediction fidelity. Our work is a case for optimism and a cautionary tale of needing uncertainty quantification and careful validation of prediction fidelity.
\end{abstract}


\bigskip
\noindent\textbf{Impact Statement}

\noindent Due to the increasing availability of global satellite ocean color data and the growing need for spatially explicit ecological habitat information, we have created a workflow addressing the following:
\begin{itemize}
\item We identify modeled plankton eco-provinces (ocean biomes) using an advanced manifold methodology allowing for uncertainty quantification.
\item Predict eco-provinces based on modeled ocean color fields using a network structure designed for fidelity verification.
\item Reveal a nuanced relationship between input data and predictability.
\end{itemize}
\noindent Our work serves as a proof of concept for predicting plankton ecosystems based on satellite ocean color data. Case study examples demonstrate the usefulness of explainability and interpretability in our workflow, highlighting it as a stepping stone for potential use in real-world applications.

\bigskip

\section{Introduction}
As the impacts of climate change become more severe, there is an increasing need for a comprehensive characterization to represent the ecology of the global ocean. Due to the shortage of global ocean in situ data, eco-provinces will need to be objectively discovered, then inferable, or predictable, from remote data. We use an unsupervised machine learning (ML) workflow to determine eco-provinces based on modeled phytoplankton functional type (PFT) concentrations, and a supervised dense network ensemble (DNE) approach to infer eco-provinces based on modeled ocean color input data. Both delineation and inference of the eco-provinces utilize novel ML methodologies that allow for trustworthiness through quantification of uncertainty, validation of fidelity, and explainability of predictions. 

Our work presents PFT eco-provinces that offer a comprehensive representation of the global ocean and a methodology for inferring them based on color data. When designing inference tools, a question that is pertinent with new remote sensing products is: What input data are needed to make accurate predictions? Using a fidelity verification methodology \citep{Suri2026} for inference, we demonstrate, using a hierarchy of ML models, what modeled ocean color inputs led to increases or decreases in skill. Our inference framework is a useful and trustworthy tool because it allows for prediction of ecological meaning based on remotely inferable input, combined with a prediction uncertainty measure and explanation of what inputs influenced correct and incorrect predictions. Our work offers a much-needed methodological advancement in identification and prediction of ecologically meaningful ocean provinces, but future work is needed before using it in real-world applications. 

In order to explore how our eco-province identification and prediction workflow could be useful as a preliminary foundation for future applications to operational efforts, we highlight the usefulness of uncertainty quantification, interpretability, and fidelity verification in our workflow by considering hypothetical examples focusing on information of potential interest for fisheries management and carbon sequestration. Since diatoms, a type of PFT, are important for food web and energy transport \citep{BBres2022, Tréguer2017b}, our first example considers that high diatom eco-provinces may be of downstream interest to fishery managers. Similarly, due to the role of coccolithophores in carbon sequestration due to their calcium carbonate shells, we perceive the potential of high coccolithophore regions as being of interest to carbon sequestration efforts. A better understanding of carbon sequestration is pertinent, due to its role in mitigating climate change \citep{USGS2008CarbonSequestration}. We also address examples of incorrect predictions,  one with high uncertainty and one with low uncertainty. These examples highlight the value of our methodology while offering practical lessons for extending to real-world applications. 

\subsection{Ocean Province Characterizations}

Previous ocean province characterizations have been based on examination of near-surface chlorophyll fields and expert knowledge \citep{Longhurst1995}, taxonomic configurations and evolutionary patterns \citep{Spalding2007}, taxonomic similarities and oceanographic processes \citep{Spalding2012}, and marine plant and animal distributions \citep{Costello2017}. Due to the large and highly complicated input data required for global ocean characterizations, more recent approaches have used ML in order to parse the input in a more data-driven manner. For example, \citet{Sayre2017} used k-means clustering on species distribution, temperature, salinity, and nutrient data. \citet{Kavanaugh2014} used a self-organizing map (SOM) and hierarchical agglomerative clustering to identify “seascape” provinces in the global ocean based on satellite-derived ocean color products, sea surface temperature, and chlorophyll a. \citet{Kavanaugh2014} seascapes are incorporated into the NOAA Coastwatch platform, illustrating the utility of such predictions for fisheries management. Our work stands in complement to the \citet{Kavanaugh2014} seascapes, but explores identification based on PFTs using a unique and highly advanced methodology explained below. We chose PFT composition for our regionalization, because it represents the influence of multiple environmental and biogeochemical factors, such as temperature, light, and nutrients, thus providing a meaningful representation of the marine ecosystem. 

A previous characterization that utilizes PFTs as an input is \citet{Sonnewald2020} who identified eco-provinces based on detailed ecological model output, probabilistic projection, density based clustering and graphs. Other phytoplankton-based global ocean characterizations include \citet{HofmannElizondo2021} who used a SOM and hierarchical clustering to characterize the global ocean based on phytoplankton species data, \citet{Kaneko2023} who used a co-occurrence ecological network of plankton taxa to identify six phytoplankton community types based on in situ rDNA metabarcoding, then predicted them using a support vector machine (SVM) based on satellite ocean color and temperature products.  Most recently, \citet{ElHourany2024} used a SOM and agglomerative clustering to identify spatiotemporally varying phytoplankton biomes based on satellite ocean color, chlorophyll-a, and an in situ gene dataset that reveals phytoplankton relative cell abundance. These ocean biomes reveal more detail in the sparse regions where the in situ gene dataset is located, so much of the open ocean is characterized as the same biome. Our work complements \citet{ElHourany2024}, in that it determines global PFT-based eco-provinces with higher complexity in pelagic regions.

Validation in the general field of geosciences, and in particular global ocean ecological characterizations using ML, is extremely challenging. This difficulty to validate is due to the sparsity of observational data and highly complex covariance structures within both modeled and available observational data. Our approach utilizes the Native Emergent Manifold Interrogation (NEMI) Method \citep{Sonnewald2023}, which is a clustering workflow developed for use in the field of earth sciences to determine regions of highly complicated, large, and non-linear data. Numerous related methods such as PCA and K-means require linearity of data, and do not allow for intuitive validation. Described fully in the methods  2.2, NEMI carefully accounts for nonlinear interactions in the input, quantifies uncertainty of eco-province identification using entropy, and allows for extensive external validation. This compatibility with nonlinear data, uncertainty quantification, and ability to validate make NEMI the best choice compared to other methods for eco-province identification. Validation in the context of geospatial unsupervised learning is extremely important, and different internal validation metrics can be misleading \citep{Jenniges2025}. NEMI allows for rigorous validation, using external metrics, and facilitates parameter selection by allowing the user to compare embeddings with different parameters in the 3D reduced dimension space. Our workflow allows us to choose any number of eco-provinces based on the desired application. For this paper, we consider the example of 10 eco-provinces as a starting point before considering higher complexity cases (i.e. a higher number of eco-provinces). 10 is the lowest number of eco-provinces that allows for sufficient complexity in eco-provinces, offering multiple levels of distinction between upwelling, oligotrophic, arctic, and southern ocean regions for example.  

The contributions of our identified eco-provinces include having an associated uncertainty quantification and being constructed based on modeled PFT concentrations, which can reveal more ecological information than satellite fields or sparse in situ data. Furthermore, we selected our eco-province methodology \citep{Sonnewald2023} to preserve both local and global patterns in the input data, in order to identify the most ecologically meaningful provinces. As a result of these input data and methodology selections, our identified eco-provinces reveal more complexity in the open ocean compared to past characterizations, and complement \citet{Sonnewald2020}, with increased spatial resolution, focus on phytoplankton and consideration of seasonal variability.

\subsection{Prediction of Eco-Provinces Based on Remotely Sensed Input Data}

Our work aims to establish an initial proof of concept for predicting eco-provinces based on satellite ocean color data. In our work, we use modeled input data to identify what data, and how much, is useful for eco-province inference. This work paves the way for using remotely sensed satellite data such as NASA's Plankton, Aerosol, Cloud and ocean Ecosystem (PACE) satellite for example, which includes ocean color, chlorophyll, and phytoplankton products. 

A goal of this paper is to lay foundations and provide a proof of concept for remotely inferring eco-provinces based on real-world remotely sensed data, such as from ocean color satellites.  As a test case, we use PFT biomass from a computer simulation (see methods section 2.1) as input to characterize eco-provinces, and modeled ocean color data to explore how well remote products can predict those provinces. PFT composition encodes the ecosystem response to several biogeochemical drivers. Defining eco-provinces based on PFT concentrations yields biologically meaningful regions while avoiding circularity when subsequently evaluating whether satellite observations can predict those provinces.   

Inferring ecosystems remotely remains a formidable challenge. We demonstrate that in our case this remote inference can be successfully accomplished, and provide a detailed analysis of the various tradeoffs associated with the number of inputs needed. Varying levels of inputs are considered in three versions of the prediction workflow. In order to predict eco-provinces based on remotely sensed fields, we explore the use of deep learning first in terms of what machinery is required for optimal network skill, a robust uncertainty quantification, and transparency in what inputs influenced a particular prediction to ensure fidelity of predictions, following the framework in \citet{Suri2026}. We optimize the predictive skill and uncertainty quantification by creating an ensemble of neural networks all dedicated to the same task (see methods section 2.3), then average their results to increase robustness following \citet{Clare2022, Sonnewald2021}. To prioritize transparency in network decisions, we use eXplainable Artificial Intelligence (XAI, see methods section 2.4), described in \citet{Arrieta2020, Mersha2024}. XAI is a set of techniques used to make AI models more transparent, allowing for user understanding of network decisions and addressing the concern of neural network inference methodology acting as a “black box.” This additional step allows the user to understand what network inputs increased or decreased the probability of correct or incorrect predictions, supports trustworthy decision making, and is necessary when using neural networks to make high-stakes management decisions. In addition, we tested what type and how much information is needed for our dense network ensemble (DNE) to function by considering which inputs influenced correct or incorrect predictions (see methods section 2.5). We consider different versions of the DNE with varying inputs with the rationale of starting simple with fewer inputs, then assessing any improvements from adding input fields.

\section{Methodology}
\subsection{Darwin Model}
This paper presents eco-province identification based on PFT concentrations estimated from the Darwin Biogeochemistry model \citet{Dutkiewicz2021} which is based on differential equations that follow the movement of matter through inorganic nutrient, living and dead organic matter and remineralization of detritus. The model also captures dynamically evolving Chl distributions, as well as the absorption and scattering of three streams of light on optically important constituents in the ocean. The model outputs irradiance reflectance which is a measure of ocean color, similar to what is captured by satellite sensors, and is divided into wavebands across the visible light spectrum. The model output spans 1993-2011, and in this study we consider phytoplankton biomass, chlorophyll, and irradiance reflectance values for the 425 nm (blue reflectance) and 550 nm (green reflectance) wavebands. Phytoplankton growth is parameterized as dependent on temperature, irradiance, and nutrients. The model captures 50 types of plankton that differ in terms of trophic strategy, functional roles and size (ranging from 0.6 to over 1000 micrometers in equivalent spherical diameter). Here we focus on groups of phytoplankton known as ``phytoplankton functional types" (PFTs): diatoms (that require silica), coccolithophores (that calcify), mixotrophic dinoflagellates (which graze as well as photosynthesize), picoeukaryotes, and picoprokaryotes (which are well adapted to low nutrient conditions). 

\begin{figure}[H]
\centering
\includegraphics[
  width=\textwidth,
  height=0.75\textheight,
  keepaspectratio
  ]{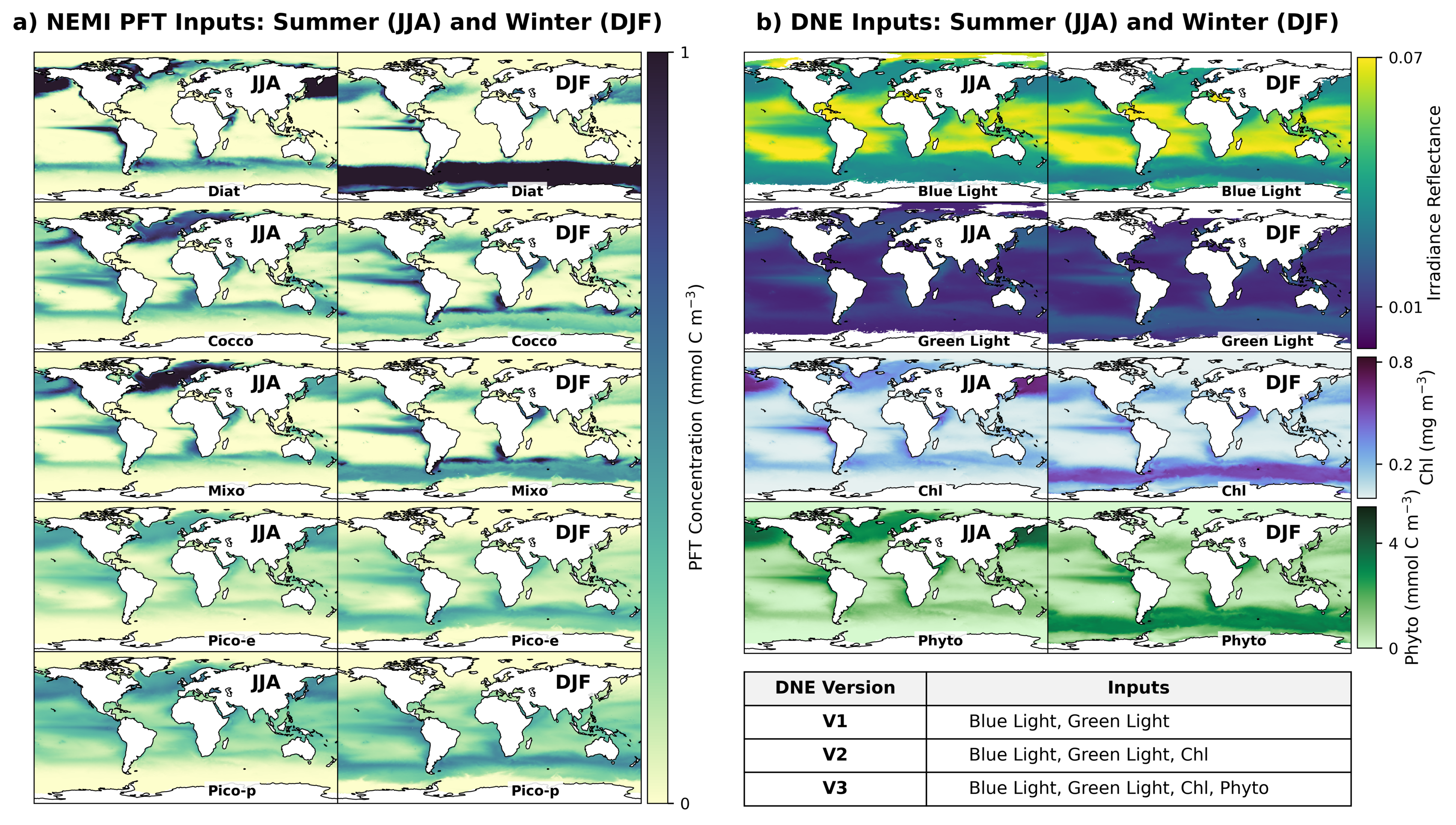}
\caption{\textbf{NEMI PFT Inputs and DNE Inputs}. Panel a) displays in the left column boreal summer (June, July, August: JJA) and in the right column boreal winter (December, January, February: DJF) NEMI PFT concentrations. The PFTs are Diatoms (Diat), Coccolithophores (Cocco), Mixotrophic Dinoflagellates (Mixo), Picoeukaryotes (Pico-e), and Picoprokaryotes (Pico-p). Panel b) presents boreal summer (June, July, August: JJA) and boreal winter (December, January, February: DJF) DNE inputs. Included is irradiance reflectance for wavebands 425 nm (blue reflectance), 550nm (green reflectance), chlorophyll (chl), and phytoplankton carbon biomass (phyto). The bottom right table indicates which inputs are included in each version of the DNE. }
\label{fig:1}
\end{figure}

All model output used here consists of 3-day averaged values on a 1/2 by 1/2 degree (about 50km) spatial grid. Here we only use the surface (0-10m) values, though the model includes 50 levels ranging in resolution of 10m at the surface to 100m at depth. To analyze general seasonal changes, we considered JJA and DJF averaged datasets across the time period of the dataset. The model captures the higher concentration of phytoplankton carbon biomass in northern latitudes JJA, compared to southern latitudes in DJF due to warmer weather and upwelling supporting phytoplankton blooms. Fig.~\ref{fig:1} presents a) diatoms thrive in the North Pacific in JJA and Southern Ocean in DJF, but are in lower abundance in the oligotrophic regions due to their higher nutrient and silica requirements. Coccolithophores are most abundant in the North Atlantic in JJA and mid to high southern latitudes in DJF. Mixotrophic dinoflagellates are abundant in mid- to high northern latitudes in JJA and mid- to high southern latitudes in DJF. Picoeukaryotes and picoprokaryotes follow similar seasonal patterns, with greater persistence in oligotrophic waters due to their small size and higher nutrient affinity. We decided to identify eco-provinces based on just PFTs to avoid redundancy, since PFTs are modeled based on nutrients and zooplankton are modeled based on PFTs.

\subsection{Eco-Province identification using NEMI}
Eco-provinces are identified using the “Native Emergent Manifold Interrogation” (NEMI) method, developed by \citet{Sonnewald2023} to determine regions of interest in large or highly complicated and nonlinear input data. The first step of NEMI is UMAP, “Uniform Manifold Approximation and Projection for Dimension Reduction,” \citep{McInnes2018} a general non-linear dimensionality reduction technique that constructs a simplified, three-dimensional representation of the five dimensional input called an “embedding.” Three dimensions were chosen to easily visualize the data. Each of the original dimensions of the input dataset corresponds to each of the five considered phytoplankton functional types. The embedding is important, because it offers a three dimensional representation, where each point in the embedding represents one point in geographical space. The UMAP dimension reduction technique is sophisticated in its minimization of categorical cross entropy between the high and low dimensional spaces, resulting in a locally and globally balanced embedding \citep{McInnes2018}. On this simplified representation, NEMI uses hierarchical agglomerative clustering to identify meaningful areas of interest in the low-dimensional representation. Step two of NEMI addresses the sensitivity of clustering results to noise and uncertainty through the use of an ensemble methodology. Each iteration, which we define as one singular run of the NEMI dimension reduction and clustering workflow with the same parameters, is repeated several times in order to create a group or “ensemble” of results. Each ensemble member has its own cluster assignments for each original data point. Entropy quantifies the uncertainty of whether each original input point belongs to the assigned cluster. In information theory, the entropy of a random variable measures the uncertainty of the variable’s potential states. For each sample $i$, the entropy $H_i = -\sum_{j=1}^{N_I}p_{ij}log(p_{ij}) $, where $N_I$ is the number of possible outcomes and $p_{ij}$ is the probability that outcome $j$ occurs for each sample \citep{Goodfellow2016}. In the case of NEMI, we are determining entropy for each datapoint $i$ by comparing the cluster assignments across each member of the ensemble. Therefore $p_{ij}$ is the proportion of times point $i$ is assigned to cluster $j$ across the ensemble. The final eco-provinces are chosen based on the ensemble member that has the least entropy. Thus each point in the global ocean is mapped to a specific eco-province, alongside a quantitative measure of uncertainty.

Unlike commonly used dimension reduction and clustering techniques such as Principal Component Analysis and k-Means, NEMI does not have strict assumptions about the linearity or spherical clusterability of the underlying input data. NEMI performs well with large and highly elaborate and nonlinear data, and is less parametric compared to other methods. Compared to \citet{Sonnewald2020}, NEMI has fewer parameters to tune and therefore less risk of noise interference. Due to the straightforward nature of the algorithm, NEMI is intuitive to use for global applications \citep{Sonnewald2023}. Since NEMI allows the user to set the number of clusters, a desirable level of complexity can be achieved for the application of interest.

\subsection{DNE for Eco-Province Inference Based on Ocean Color Data}

After eco-province identification, we assess prediction of eco-provinces from remotely sensed products. We use a dense ensemble multilayer perceptron (MLP) on Darwin Model output that is similar to products from satellite sensors (reflectance irradiance, total chlorophyll and total phytoplankton carbon biomass). There is some stochasticity involved in training a neural network due to random weight initialization, random mini-batch (subset) sampling, and Adam optimization which is an extension of stochastic gradient descent. As a result, independently trained networks may converge to different local optima and produce slightly different predictions, even when the same architecture and training data are used. In addition, there is a risk of overfitting or becoming stuck in local minima of the loss function. To address this stochasticity, we use a group, “ensemble,” of 10 neural networks dedicated to the same prediction task. Each neural network uses categorical cross entropy for the loss function. The final prediction is a mean of the ensemble’s results, which is much more robust than a prediction from a single neural network because it averages out randomness, reduces overfitting, and allows uncertainty estimation. If all ensemble members agree on the prediction, the final result is more reliable. Averaged ensemble entropy quantifies the uncertainty of the ensemble’s results. Our code for ensemble prediction and uncertainty quantification follows that presented in \citet{Yik2023}. We further assess the impact on ensembling using XAI, as described below. The combination of using a network ensemble for inference and XAI for interpretation increases the fidelity of our workflow \citep{Suri2026}. 

We utilize three versions of the DNE in order to consider varying levels of uncertainty in our input fields. A summary of these versions is included at the bottom of Fig.~\ref{fig:1}b. V1, which has the least input uncertainty, only considers blue and green irradiance reflectance (ocean color) inputs. There is more blue reflectance in the oligotrophic gyres due to the low abundance of phytoplankton carbon biomass in that region. In contrast, there is more green reflectance in high phytoplankton biomass regions such as upwelling regions mid to high northern latitudes in JJA, and Southern Ocean in DJF. V2 considers ocean color and total chlorophyll measured in $\frac{\text{mg Chl}}{m^3}$. Chlorophyll follows the seasonal pattern noted previously. V3 includes the addition of total phytoplankton biomass measured in $\frac{\text{mmol Carbon}}{m^3}$. Although there are some differences, phytoplankton carbon biomass follows similar seasonal patterns. 

In order to train the DNE, we first split the input data into training, test, and validation sets by splitting up longitudinal regions on the globe (Fig. S2) following \citep{Yik2023}. For an additional level of validation of the DNE in addition to prediction on the vertical test region, we predict the 2005 eco-provinces for the entire globe. Each version of the model is associated with a quantified uncertainty, computed across the ensemble. 

\subsection{SHapley Additive exPlanations (SHAP)}
Finally, SHAP, \citep{Lundberg2017}, was employed to identify the influence of particular features on where the DNE correctly and incorrectly predicted 2005 eco-provinces, based on the labels for the 1993-2004 dataset. SHAP takes a game theoretic approach, testing every combination of inputs in order to assess the contribution of each individual input. High SHAP values indicate that the input variable increased the probability of the prediction, while low SHAP values reveal the feature decreased the probability of the prediction. In our analysis, we distinguished SHAP values in locations where the neural network made correct vs. incorrect predictions, in order to more accurately pinpoint input influence.

\subsection{Advantages of Our Methodology}
Our methodology considers optimal processes throughout both the eco-province identification and prediction stage which is important and necessary based on the intricate nature and large size of our input data. In the identification phase, NEMI uses UMAP which is a topological dimension reduction technique that involves the use of categorical cross entropy to optimize representation of both high and low dimensional space in the embedding which preserves structure well for large and highly complicated datasets, as demonstrated for marine geospatial data in \citet{Jenniges2025}. The agglomerative clustering approach allows the user to select the number of clusters, which makes our approach more customizable to the desired application. In addition, our NEMI eco-provinces have a mathematical uncertainty measure for each grid cell. Our DNE construction is much more robust compared to just a single neural network prediction. The ensemble prediction includes prediction entropy averaged across the ensemble, which allows for a more transparent and interpretable inference tool. Furthermore, three versions of our DNE allow us to identify the importance of the number of inputs for prediction accuracy in different locations across the globe. SHAP adds another layer of transparency to the predictions, assigning a measure of importance to each input towards correct or incorrect predictions of each class. \citet{Suri2026} emphasize the importance of such a workflow for verifying the fidelity of learned physical dependencies to known dynamics. The ability to predict verified ocean community structure based on remotely sensed input data, alongside an uncertainty measure, is very important and timely as a baseline to monitor climate change.

\subsection{DNE \& XAI Workflow Overview}
\begin{figure}[H]
\centering
\includegraphics[width=\linewidth]{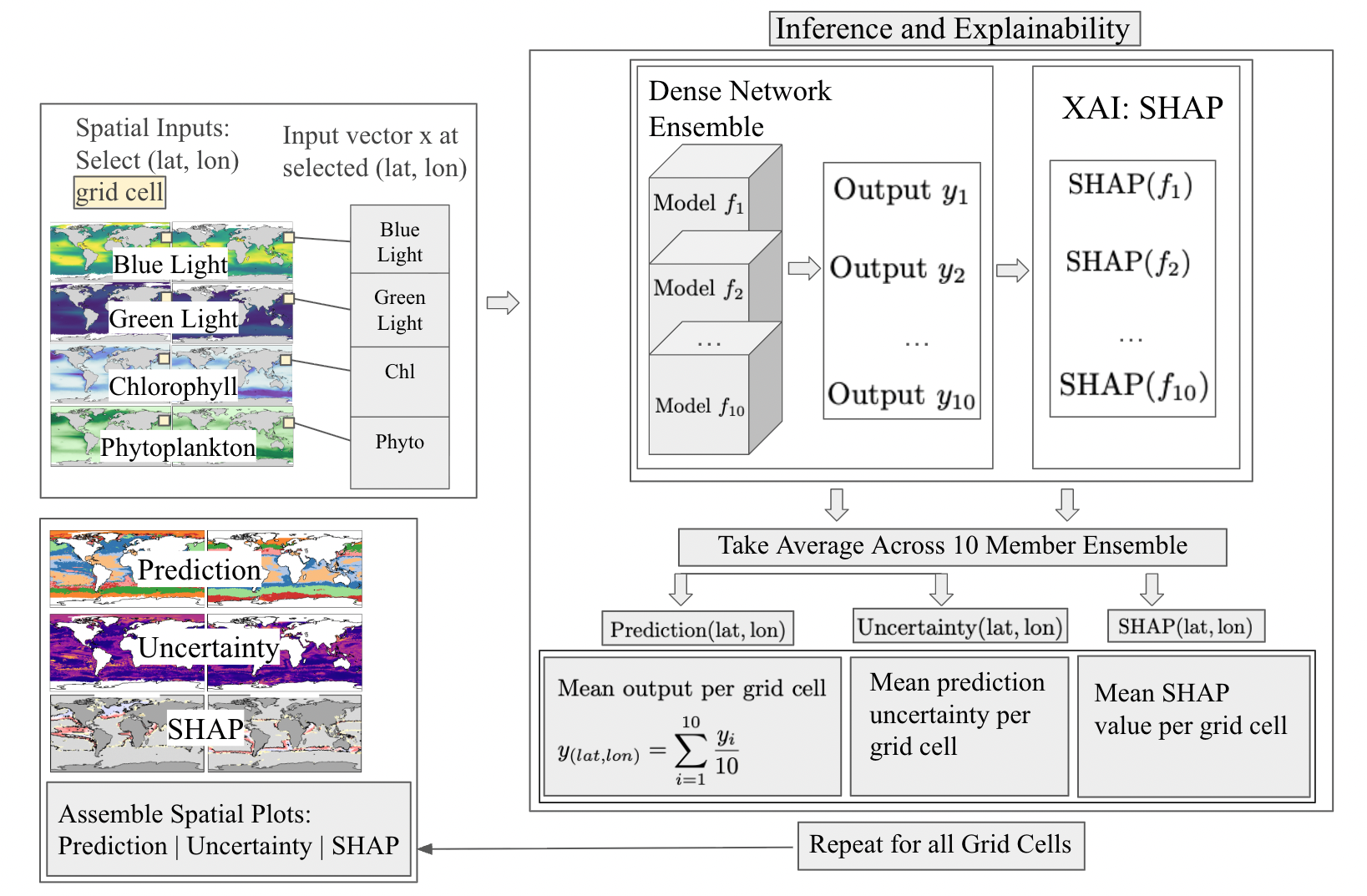}
\caption{\textbf{DNE Construction}. The construction of the DNE, including the modeled input data that can be remotely sensed, a grid cell by grid cell input into the robust DNE prediction, repetition for all grid cells, average prediction and entropy across the ensemble, and SHAP to determine the importance of each input for correct and incorrect grid cell predictions is displayed.}
\label{fig:3}
\end{figure}

Fig.~\ref{fig:3} presents the neural network construction for all three versions predicting a single year along with an uncertainty quantification, and the workflow for using SHAP to assess the contribution of each input to correct and incorrect predictions.

\section{Results}

 Our ML inputs are from the Darwin Biogeochemistry Model \citep{Dutkiewicz2021} which is a marine biogeochemical-ecosystem numerical model that includes an explicit radiative transfer module, and hence is able to provide reflected irradiances (similar to what a satellite might sense), nutrients, chlorophyll, and over 50 types of phytoplankton which we group into 5 PFTs, picoprokaryotes, pico-eukaryotes, coccolithophores, diatoms and mixotrophic dinoflagellates. We use PFTs for eco-province identification (Fig.~\ref{fig:1}a) and input similar to what satellite sensors estimate (blue reflectance, green reflectance, chlorophyll, and phytoplankton carbon biomass) for eco-province prediction (Fig.~\ref{fig:1}b). Eco-province identification uses the NEMI clustering algorithm, based on boreal summer (June, July, August: JJA) and boreal winter (December, January, February: DJF) phytoplankton functional type biomass estimates averaged across 1993-2015.

\subsection{Seasonal Eco-Province Identification}

\begin{figure}[H]
\centering
\includegraphics[
  width=\linewidth,
  height=0.75\textheight,
  keepaspectratio
]{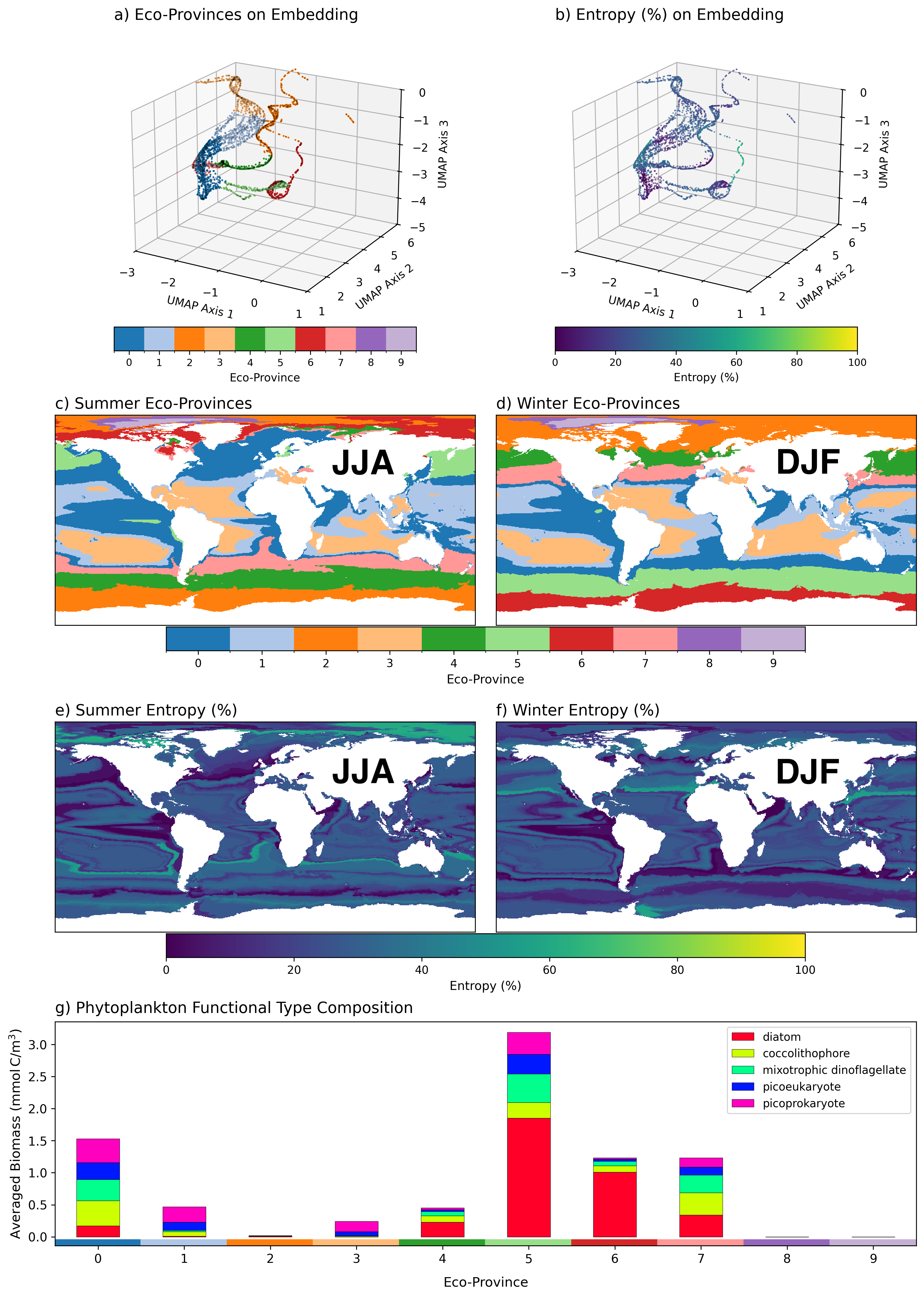}
\caption{\textbf{Embedded and spatial Eco-Provinces With Corresponding Uncertainty, and PFT Input Composition}. 
Section a) presents the 10 JJA boreal summer and DJF boreal winter eco-provinces plotted on the embedding, which is produced by UMAP dimension reduction. The corresponding entropy (uncertainty) is plotted in b). Sections c) and d) display the 10 JJA and DJF eco-provinces plotted geographically, with the corresponding uncertainty measured with entropy in e) and f). g) displays the phytoplankton functional type composition for each eco-province.
}
\label{fig:2}
\end{figure}

Fig.~\ref{fig:2}a illustrates the NEMI eco-provinces on the embedding. Each point corresponds to a spatial grid cell (latitude and longitude) on the JJA or DJF plot, and the color represents the eco-province label. The embedding offers a visualization of the data in three dimensional space, which is important for external validation of the eco-provinces. The eco-province assignment for each grid cell corresponds to an uncertainty percentage quantified by entropy, presented in Fig.~\ref{fig:2}b. Entropy quantifies the uncertainty of each cluster assignment for each grid cell through comparison of cluster assignments across an ensemble, which allows the final eco-provinces to be chosen based on the ensemble member with the least uncertainty across the globe. As the color bar indicates, dark blue regions represent eco-provinces with lower uncertainty compared to lighter blue and green regions. Here, we chose the number of eco-provinces to be 10, but any number could be chosen based on the desired application. Note that the number of points presented in Fig.~\ref{fig:2}a and Fig.~\ref{fig:2}b is randomly subsampled for visualization.

As each point in Fig.~\ref{fig:2}a represents one location, we can project these from the embedding space back onto geographical space. Fig.~\ref{fig:2}c and Fig.~\ref{fig:2}d present the eco-provinces plotted geospatially, for averaged JJA (Fig.~\ref{fig:2}c) and boreal winter DJF (Fig.~\ref{fig:2}d). The same eco-provinces are identified in both JJA and DJF highlighting a seasonal shift in eco-province location, especially in higher latitudes. Although expected, the nuances in eco-province grouping cannot be pre-determined without the unsupervised NEMI methodology. For example, just viewing the phytoplankton carbon or chlorophyll biomass in Fig.~\ref{fig:1}b would not indicate the nuances in the Arctic in eco-provinces 8 (dark purple) and 9 (light purple), or the split of eco-province 2 between northern and southern latitudes in JJA. 

While eco-provinces in the lower latitudes remain more consistent across the seasons, eco-provinces in high latitudes are more variable. Since eco-province 3 (light orange) exists in low nutrient regions, we will refer to it as the oligotrophic eco-province. Similarly, since eco-province 1 (light blue) serves as a thin transition between two larger eco-provinces, we will refer to it as the border eco-province. Eco-provinces 8 (dark purple) and 9 (light purple) are located in the Arctic, so we will refer to them as the Arctic provinces. These eco-provinces (oligotrophic, border, and Arctic) do not change very noticeably depending on the season. In contrast, eco-provinces 2 (dark orange), 4 (dark green), 5 (light green), 6 (red), and 7 (salmon) shift from high southern to northern (or vice versa) depending on the season. Eco-province 2 characterizes high latitudes, while eco-province 5 encompasses regions with high diatom biomass. Note that eco-provinces 0 (encompassing coastal and upwelling regions, dark blue), 5 (light green) and 6 (red), for the most part tend towards northern latitudes in JJA and southern latitudes in DJF. In contrast, eco-provinces 2, 4, and 7 exhibit the opposite pattern, tending towards southern latitudes in JJA. 

Fig.~\ref{fig:2}e and Fig.~\ref{fig:2}f show the JJA and DJF spatial entropy percentages, determined by the cross-ensemble entropy described above. Note that lighter blue and green regions have higher uncertainty than dark blue regions. For example, eco-provinces 1 and 6 have more uncertainty than eco-province 3. In general, the uncertainty is higher between clusters, which is expected. The core regions of each cluster are less variable, but the borders vary slightly on each run of the algorithm. This uncertainty quantification is important since it allows us to map quantified entropy to each point in the ocean to its specific eco-province. Note that since the cluster assignments and corresponding uncertainty quantifications are for each grid cell individually, the certainty can vary across one cluster. In general, high uncertainty regions are those where the DNE struggles to make predictions.

We compare the clustering results with the original modeled PFT biomass (Fig.~\ref{fig:2}g) in order to gauge the ecological fidelity of the results. There are more diatoms, mixotrophic dinoflagellates, picoeukaryotes, and picoprokaryotes in cluster 5 and more coccolithophores and picoprokaryotes in cluster 0. The PFTs most abundant in cluster 5 have a strong seasonal component where they exist in the mid to high northern latitudes in JJA, then mid to high southern latitudes in DJF. The PFTs most abundant in cluster 0 exist surrounding continents and oligotrophic gyres, with a slight seasonal polarity that emphasizes a strong presence in the North Atlantic and Pacific during JJA. There is nearly no phytoplankton in eco-provinces 8 and 9 due to ice in those regions. 

The transition zones between eco-provinces have higher uncertainty compared to larger, core eco-provinces. Some general eco-province structures can be explained by physics-based knowledge. For example, the Southern Ocean is broken up into two provinces during both JJA and DJF, due to ACC (Figs.~\ref{fig:2}c,d). The oligotrophic gyres, which are strongly influenced by global wind patterns, are characterized by eco-province 3 in both JJA and DJF. General province structures, most noticeably the horizontal v shape surrounding the Eastern Tropical Pacific, align with large-scale “planetary waves” caused by Earth’s rotation. However, the eco-provinces cannot be expected from physical knowledge alone. For example, the location of the low nutrient oligotrophic eco-province does not align with expected gyre structure in the North Atlantic.



\subsection{Seasonal Eco-Province Inference}

Having defined the eco-provinces from PFT biomass input, we now consider how well remotely sensed products could be used to infer the provinces. Our work considers three versions (V1, V2, and V3) of the DNE, with different input features as specified in Fig.~\ref{fig:1}b. V1 includes blue and green irradiance reflectance, a measure of ocean color, as input. V2 also includes total chlorophyll as an input. V3 additionally includes total phytoplankton carbon biomass. These inputs (reflectance, Chl, carbon biomass) are typical satellite products. Note that Chl and Carbon biomass are determined from the reflectance, so these fields are correlated \citep{Graff2015}. Fig.~\ref{fig:3} outlines the general inference and explainability workflow, which includes the process of DNE prediction with uncertainty and explainability for each grid cell.

\begin{figure}[H]
\centering
\includegraphics[width=\linewidth]{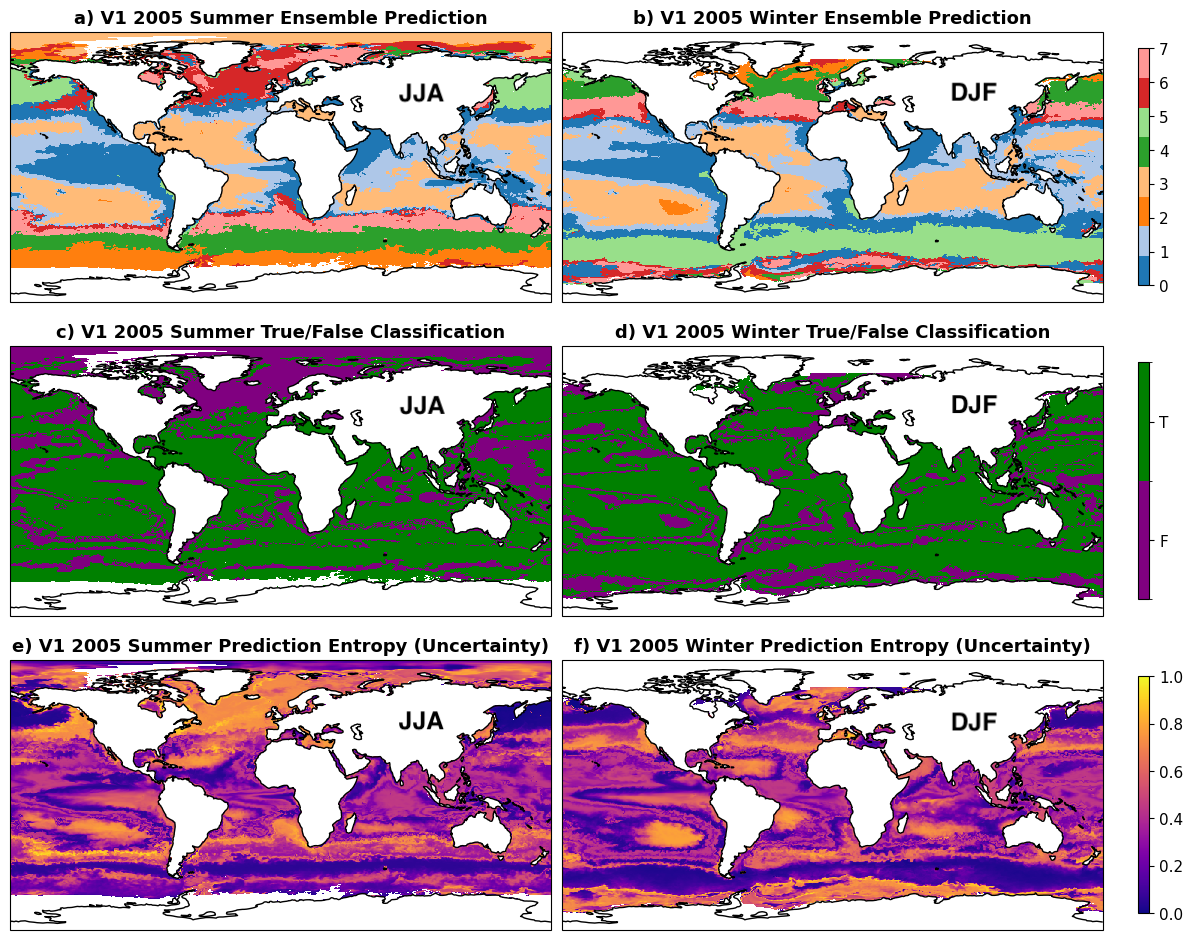}
\caption{\textbf{V1 predicted eco-provinces, with corresponding true-false and entropy plots}. Sections a) and b) illustrate the V1 averaged prediction of 2005 boreal summer (June, July August: JJA) and boreal winter (December, January, February: DJF) eco-provinces. V1 includes blue and green reflectance as DNE inputs. Sections c) and d) present where the prediction was correct in green and incorrect in purple, compared to the actual NEMI eco-province labels. Note the incorrect predictions in the Arctic, Western Pacific, and in between eco-provinces. Sections e) and f) display the averaged entropy (uncertainty) of the prediction. The incorrect prediction of the boreal summer (JJA) Western Pacific has fairly low uncertainty.}
\label{fig:4}
\end{figure}

Fig.~\ref{fig:4} displays the V1 predicted 2005 eco-provinces (after training on 1993-2004 input data) with corresponding true/false values and entropy. The true/false values indicate correct and incorrect network predictions, respectively. True values are assigned to grid cells where the DNE's prediction aligned with the identified eco-provinces, and false values are where the prediction differed. Prediction entropy, calculated from the DNE's predicted class probabilities, quantifies the DNE's classification confidence. DNE prediction entropy is distinct from eco-province identification entropy, which measures agreement in eco-province assignment across the ensemble. In Fig.~\ref{fig:4}c and Fig.~\ref{fig:4}d, the correct DNE predictions are displayed in green compared to the incorrect predictions in purple. Note the incorrect predictions in the Arctic, Western Pacific, and in between eco-provinces. Fig.~\ref{fig:4}e and Fig.~\ref{fig:4}f display the uncertainty of the prediction quantified as entropy, where darker regions have less uncertainty. We find interesting deviations from the correspondence of skill and prediction uncertainty, where the DNE was certain the prediction was correct, but it was wrong. The incorrect prediction of the boreal summer (JJA) Western Pacific has fairly low uncertainty, which is unexpected for an incorrect prediction. We discuss the low uncertainty incorrect prediction of the Western Pacific later as a case study of a cautionary tale that emphasizes the need for prediction explainability and uncertainty quantification. 

\begin{figure}[H]
\centering
\includegraphics[width=\linewidth]{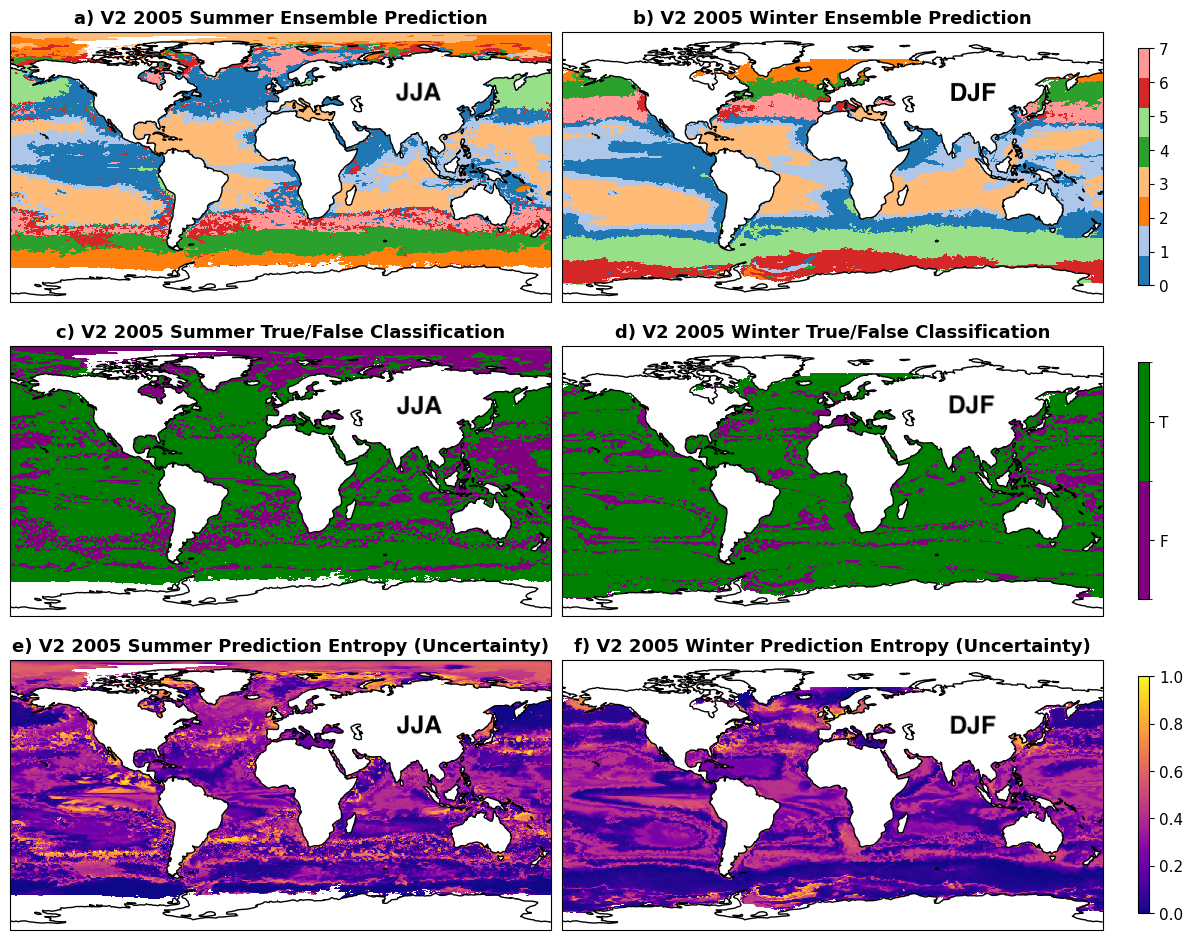}
\caption{\textbf{V2 predicted eco-provinces, with corresponding true-false and entropy plots}. Fig.~\ref{fig:5} has the same format as Fig.~\ref{fig:4}, but for V2 which has chlorophyll as a DNE input, in addition to the blue and green reflectance in V1. Boreal summer corresponds to June, July, August (JJA) and boreal winter corresponds to December, January, February (DJF). Compared to V1, the prediction improves in the North Atlantic, but gets worse in the Western Pacific and the Southern Ocean. Note the high entropy in the incorrect Southern Ocean prediction, but low entropy associated with the Western Pacific incorrect prediction.}
\label{fig:5}
\end{figure}
\begin{figure}[H]
\centering
\includegraphics[width=\linewidth]{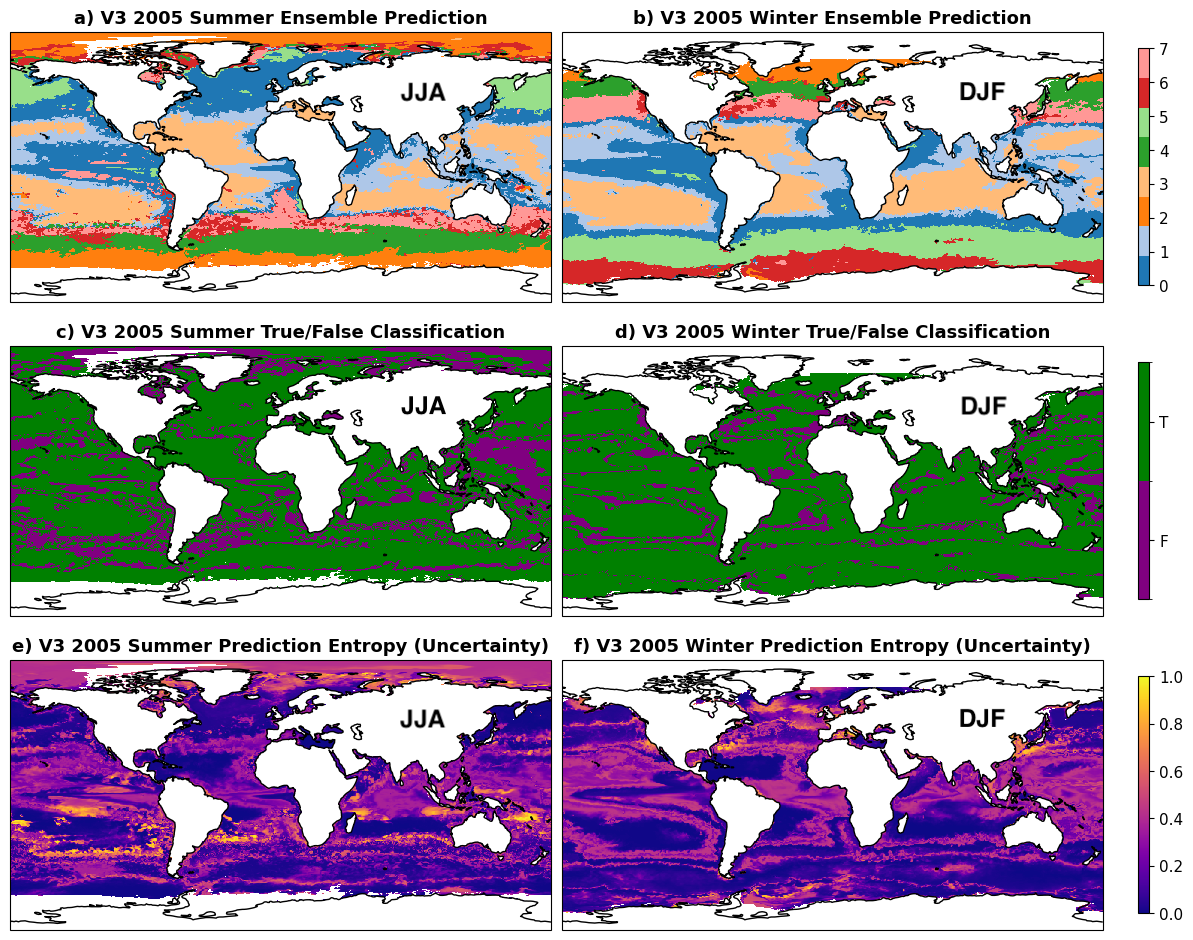}
\caption{\textbf{V3 predicted eco-provinces, with corresponding true-false and entropy plots}. Here are the boreal summer (June, July, August: JJA) and boreal winter (December, January, February: DJF) results of V3, which includes phytoplankton carbon biomass as an input in addition to the inputs in V2. Note the continued improvement in the Arctic, alongside sustained difficulty in the summer (JJA) Southern Ocean and Western Pacific. As previously observed, the uncertainty of the incorrect Southern Ocean prediction is high compared to the uncertainty of the incorrect Western Pacific prediction.}
\label{fig:6}
\end{figure}

Figs.~\ref{fig:5} and \ref{fig:6} are analogous to Fig.~\ref{fig:4}, but for network versions 2 and 3. When we add chlorophyll as an input, the prediction improves in the JJA North Atlantic, but degrades in JJA just north of the Southern Ocean and Western Pacific (Fig.~\ref{fig:4}). Explanations for why the prediction degraded in these locations are explored through the use of XAI below. The incorrect prediction just north of the Southern Ocean has high uncertainty, which is expected of an incorrect prediction. Adding phytoplankton carbon biomass as an input further improves the prediction in the JJA Arctic, but the incorrect predictions of the JJA top of the Southern Ocean and Western Pacific remain. 

The DNE predicts core ocean provinces (Table S1) such as the oligotrophic gyres, Southern Ocean (eco-province 2), and upwelling provinces with more accuracy (96\%, 96\%, 87\% respectively) compared to border provinces such as those surrounding the oligotrophic gyres and in between the Arctic and North Atlantic (78\% and 63\% respectively). Better predictability of core regions aligns with expectations, as they do not change as frequently in space and time, and were initially identified with higher certainty.


Comparing all three versions of the DNE in terms of the overall accuracy, more input variables lead to better performance, but with interesting deviations. The addition of the extra inputs (going from V1 to V2 to V3) increases the maximum training accuracy from 0.78 to 0.85 and 0.88, respectively. The most improvement is observed in the prediction of the JJA North Atlantic and JJA Arctic with increasing versions as more inputs are added to the predictive model. Across the three versions, the maximum overall validation accuracy increases from 0.67 to 0.81 and 0.85. Furthermore, the respective minimum training and validation losses decrease from 0.53 and 0.82 for V1, to 0.37 and 0.49 for V2, to 0.30 and 0.37 for V3. Overall, there are more occurrences of false predictions on the boundaries compared to the cores of each province. Accordingly, the prediction entropy is higher between provinces. The entropy distributions are lower for correctly classified grid cells compared to incorrectly classified grid cells. From a DNE performance perspective, V3 is desirable compared to the others. The DNE’s prediction accuracy increases with the addition of chlorophyll and phytoplankton carbon biomass as inputs because additional inputs have an increased variety of maxima and minima values to distinguish distinct provinces.

In our evaluation, we match skill with uncertainty, where largely a poor prediction has high uncertainty, but this is not always the case. The DNE predicts core provinces such as the oligotrophic gyres and areas in the Southern Ocean with much more accuracy compared to border provinces such as cluster 1 where it surrounds the oligotrophic gyres, across all versions. 

Our results strongly indicate that the best version of input data depends on the user's region of interest. Fig. S3 displays the 3 compositions of DNE inputs in each eco-province, so that we can determine whether high or low concentrations of each input contributed to correct (true) or incorrect (false) predictions displayed in the true-false plots. In general, either high or low input concentrations are more informative for DNE predictions compared to moderate concentrations.

\subsection{Inference Fidelity Assessment Across Three Versions of DNE}

\begin{figure}[H]
\centering
\includegraphics[width=\linewidth]{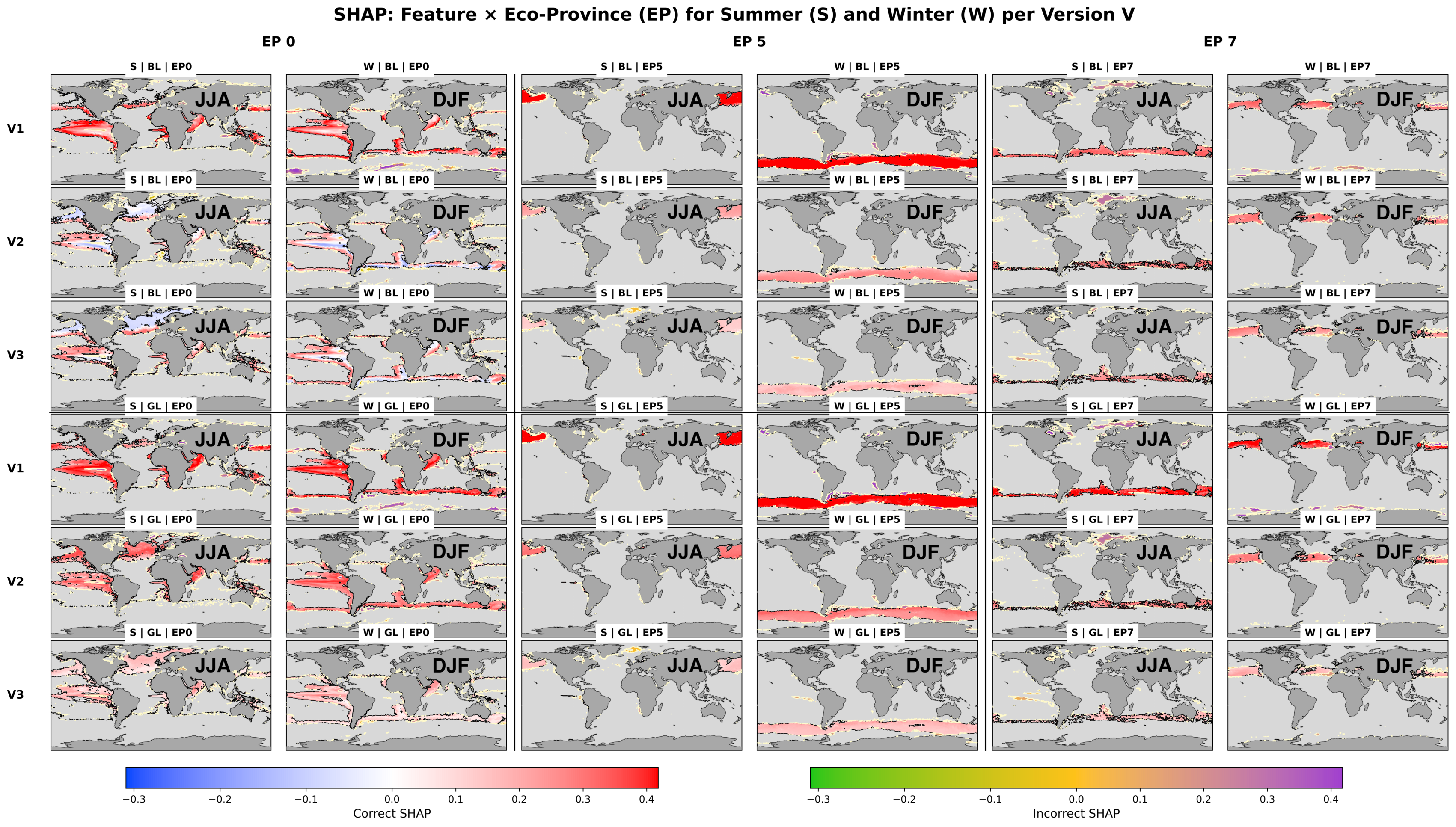}
\caption{\textbf{SHAP results for Eco-Provinces (EPs) 0, 5, and 7 for all three versions of the DNE, including blue and green reflectance inputs.}
For boreal summer (JJA) and boreal winter (DJF) the importance of blue and green reflectance decreases across versions with the addition of chlorophyll and phytoplankton carbon biomass. Blue reflectance is more important for prediction on the boundary of eco-province 0, compared to green reflectance which is more influential in the center of the eco-province. Also note the high accuracy of predictions for these core eco-provinces 0, 5, and 7, which in S3 correspond to fairly high DNE inputs. More specifically, eco-province 0 has high levels of all inputs, eco-province 5 has high levels of all inputs except for blue reflectance, and eco-province 7 has medium levels of blue reflectance and fairly high levels of the other inputs.}
\label{fig:7}
\end{figure}

\begin{figure}[H]
\centering
\includegraphics[width=\linewidth]{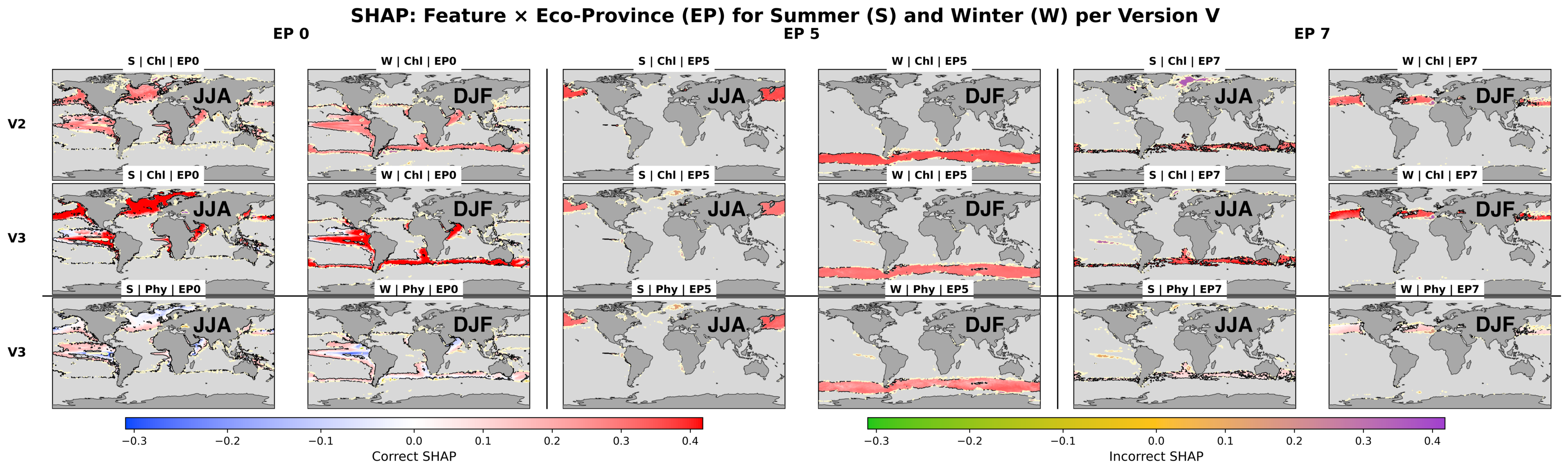}
\caption{\textbf{SHAP results for Eco-Provinces (EPs) 0, 5, and 7 for all three versions of the DNE, including chlorophyll and phytoplankton carbon biomass inputs.} Displayed are the boreal summer (June, July, August: JJA) and boreal winter (December, January, February: DJF) SHAP values for eco-provinces 0, 5 and 7 for chlorophyll and phytoplankton carbon biomass inputs across all three versions. Overall, chlorophyll is most influential in these predictions. Note that chlorophyll is more important in some cases for predicting the inside region of eco-province 0, compared to phytoplankton which is more important on the outside of the eco-province.}
\label{fig:8}
\end{figure}

\begin{figure}[H]
\centering
\includegraphics[width=\linewidth]{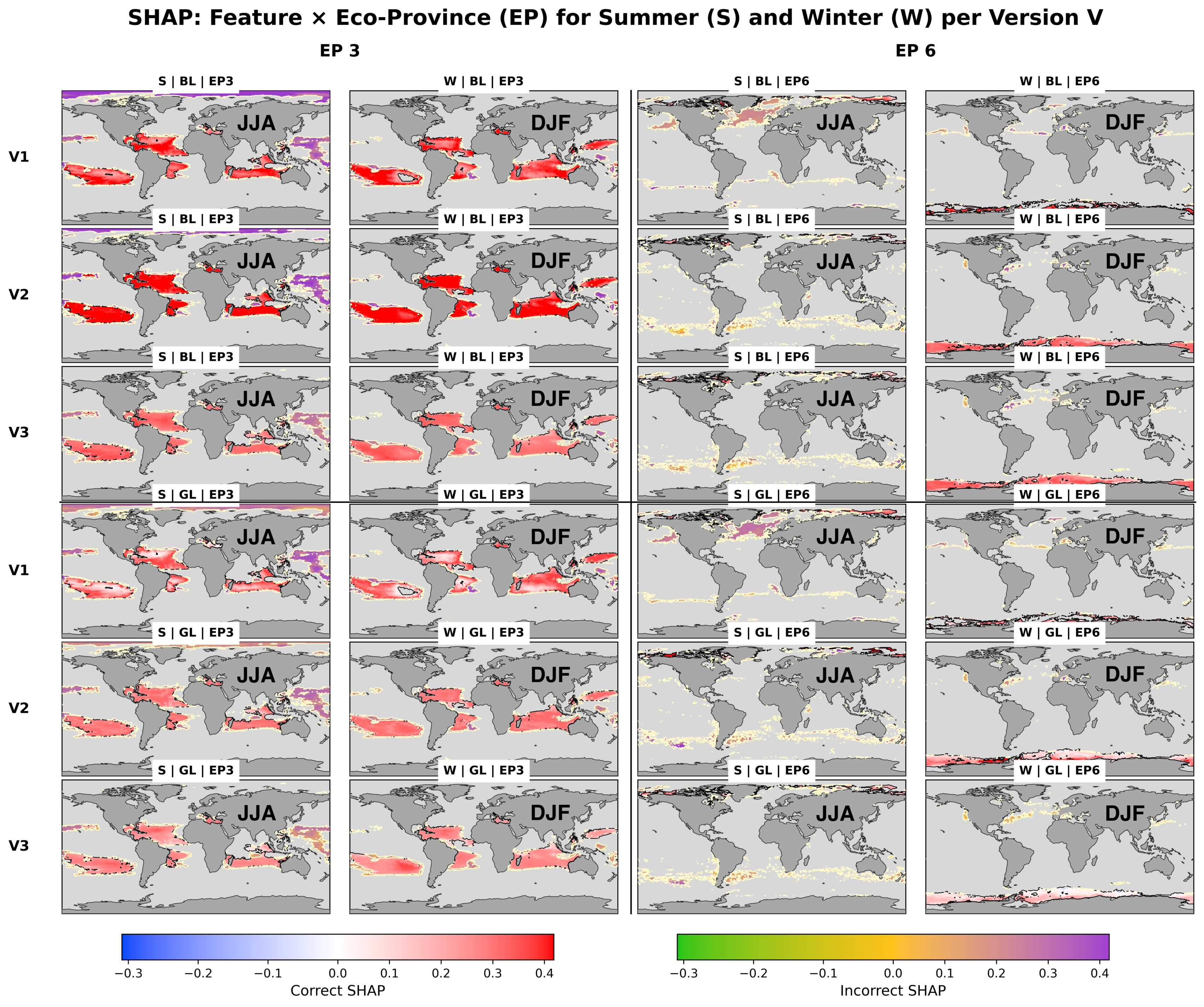}
\caption{\textbf{SHAP values for eco-provinces (EPs) 3 and 6 for versions 2 and 3, including blue and green reflectance inputs}. The importance of blue and green reflectance across all three versions for the boreal summer (June, July, August: JJA) and boreal winter (December, January, February: DJF) prediction of eco-provinces 3 and 6 is demonstrated. Both inputs contribute to the incorrect predictions of these eco-provinces. Note that the incorrect prediction of eco-province 3 in the boreal summer (JJA) Western Pacific is much more surprising compared to the incorrect prediction of eco-province 6 in the boreal summer (JJA) Southern Ocean, due to the low uncertainty of the eco-province 3 incorrect prediction.
}
\label{fig:9}
\end{figure}

\begin{figure}[H]
\centering
\includegraphics[width=\linewidth]{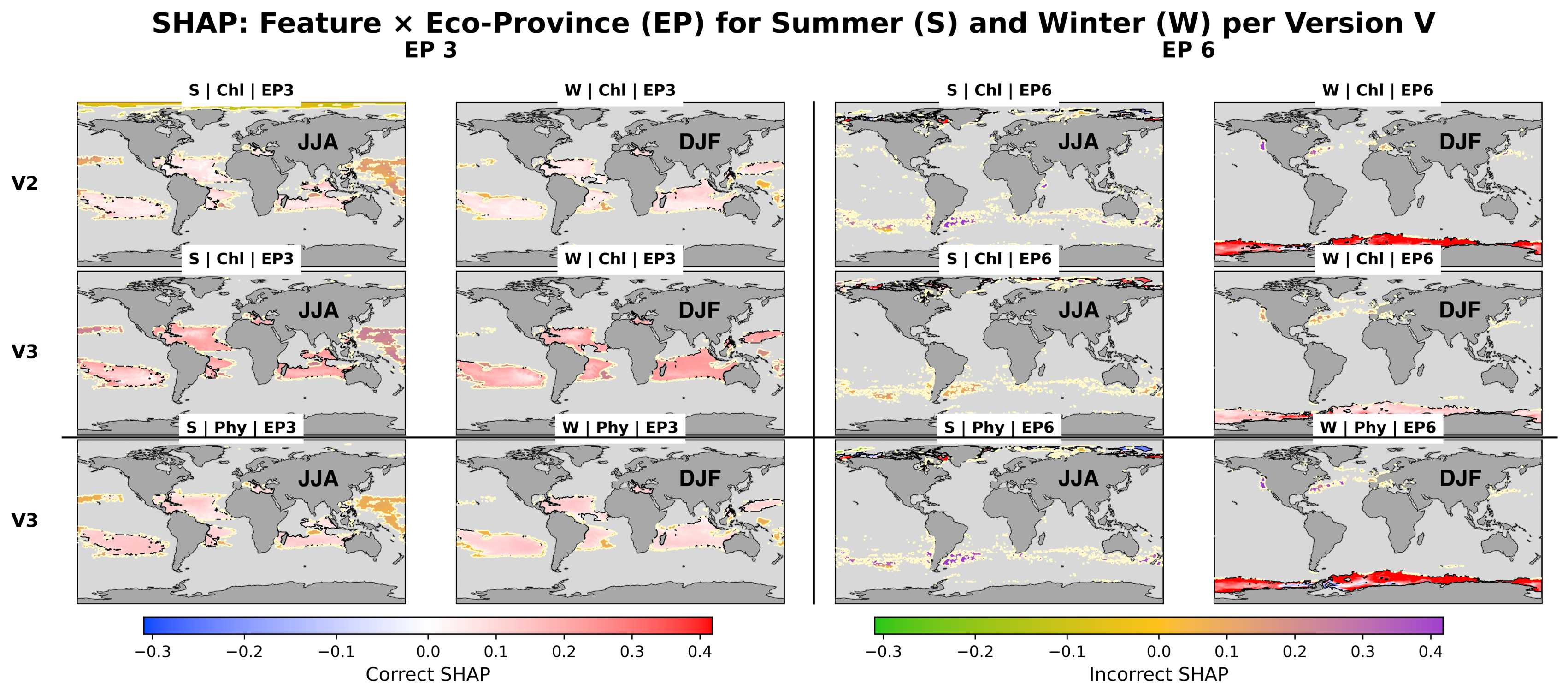}
\caption{\textbf{SHAP results for eco-provinces (EPs) 3 and 6, for chlorophyll and phytoplankton carbon biomass as inputs, across versions 2 and 3}. The importance of chlorophyll to the incorrect prediction of boreal summer (June, July, August: JJA) eco-province 3, especially in V3 which includes all four inputs, is emphasized. For eco-province 6, chlorophyll is most influential to the incorrect prediction in V2, while phytoplankton carbon biomass is most influential for V3.}
\label{fig:10}
\end{figure}

SHapley Additive exPlanations (SHAP) is an XAI technique that uses a game theoretic approach to computing the contribution of each input to a prediction \citep{Lundberg2017}. SHAP Values reveal the importance of input features (Figs.~\ref{fig:7}--\ref{fig:10}, feature importance is color coded and separated based on whether the DNE's prediction was correct or not). In the correctly predicted eco-provinces, blue represents low SHAP values where the input decreases the likelihood of the correct prediction, compared to red which indicates high SHAP values which increase the likelihood of the correct prediction. Similarly, the incorrectly predicted regions have low SHAP values in green, going through orange as neutral, and reaching purple as high SHAP values. The correctly predicted eco-provinces are outlined in black, while the incorrectly predicted regions are outlined in light yellow. We obtained the SHAP values for all eco-provinces and inputs, but for conciseness are highlighting eco-provinces 5, 0, and 7 (Figs.~\ref{fig:7},\ref{fig:8}) as examples to demonstrate the usefulness of explainability and fidelity verification in our methodology in relation to information of potential interest for future efforts focused on fisheries management or carbon sequestration. We also consider eco-provinces 3 and 6 (Figs.~\ref{fig:9},\ref{fig:10}) as case studies of incorrect predictions with low and high uncertainty, respectively.

Fig.~\ref{fig:7} displays SHAP values for blue and green reflectance, including correct and incorrect predictions plotted together, for V1, V2, and V3, for eco-provinces 0, 5, and 7.  Fig.~\ref{fig:7} reveals how the importance of blue and green reflectance decreases across versions with the addition of chlorophyll and phytoplankton carbon biomass.

Comparing the modeled ocean color inputs in Fig.~\ref{fig:7}, blue reflectance is more important for prediction on the boundary of eco-province 0, compared to green reflectance which is more influential in the center of the eco-province. Considering the addition of chlorophyll and phytoplankton carbon biomass inputs, chlorophyll is most important for predicting eco-province 0, as revealed by the SHAP values displayed in Fig.~\ref{fig:8}. Also note the high prediction accuracy of eco-province 0 corresponds to high levels of all V3 inputs (Fig. S3). Note that chlorophyll is more important in some cases for predicting the inside region of eco-province 0, compared to phytoplankton carbon biomass which is more important on the outside of the eco-province.

Considering all of the inputs, both blue and green reflectance are most influential for predicting eco-province 5 in V1 (Fig.~\ref{fig:7}) and chlorophyll is most influential in V2 and V3. The high prediction accuracy of eco-province 5 corresponds in Fig. S3 to high levels of all inputs except for blue reflectance. Green reflectance is most influential for predicting eco-province 7 in V1 (Fig.~\ref{fig:7}), while chlorophyll is most influential in V’s 2 and 3. Fig. S3 reveals that eco-province 7 has medium levels of blue reflectance and fairly high levels of the other inputs. Fig.~\ref{fig:9} displays the JJA and DJF SHAP values for eco-province 3 for versions 2 and 3, including blue and green reflectance inputs. Comparing all inputs, both blue and green reflectance, but especially blue reflectance, are influential for the incorrect prediction of the JJA Western Pacific eco-province 3, with decreasing importance across versions with the addition of chlorophyll and phytoplankton carbon biomass as inputs (Fig.~\ref{fig:9}). Figs.~\ref{fig:9} and \ref{fig:10} demonstrate the high importance of chlorophyll in the V2 and phytoplankton carbon biomass in the V3 incorrect prediction of the Southern Ocean. Fig. S3 demonstrates medium to high blue reflectance, and medium green reflectance feeding into the JJA eco-province 6 prediction.


Overall SHAP results demonstrate that the difference in prediction accuracy and uncertainty in core vs. border eco-provinces may relate to these eco-provinces containing either very high or very low input concentrations, which is more meaningful for the DNE to distinguish meaningful eco-provinces compared to less extreme input values. Overall SHAP results reveal that high blue reflectance, low green reflectance, low plankton, and low chlorophyll increase the likelihood of low plankton eco-provinces, such as the oligotrophic eco-province, being predicted. Similarly, low blue reflectance, high green reflectance, high plankton, and high chlorophyll help predict high plankton eco-provinces such as coastal upwelling provinces, the Eastern Tropical Pacific, DJF Southern Ocean, and JJA North Pacific. Furthermore, these border regions had more uncertainty during the identification phase, so the DNE having more uncertainty predicting these regions confirms expectations.


\section{Discussion}
We identify eco-provinces on a global 1/2 degree spatial grid for boreal summer (JJA) and boreal winter (DJF) averages spanning 1993-2015 from model output of phytoplankton biomass. We determine how well satellite products may be able to predict these eco-provinces by inferring them based on modeled inputs. We predict the eco-provinces using a series of DNEs, assessing how prediction accuracy and learning strategies are shaped by the inputs, training on 1993-2004 input data and then predicting 2005 global eco-provinces. To delve into the root of correct and incorrect predictions across the three versions, we use XAI to reveal input influence for case study examples. This framework, demonstrated in Fig.~\ref{fig:3}, aligns with the general workflow demonstrated in \citep{Suri2026} as a quantitative pathway to demonstrate fidelity. As we increase the number of inputs, overall accuracy increases, but the accuracy in particular regions decreases. This result aligns with expectations, because increasing input information the DNE uses has the potential benefit of increasing the context necessary for predictions, but it can also increase input data uncertainty. We challenge and test the assumption that “more input data is better,” identifying spatially which predictions improve as a result of adding more inputs. Insight from asking such questions guides our recommendations on how the DNE should be used based on different goals and in different geographical provinces.


\subsection{Identified Eco-Provinces}
The identified eco-provinces exhibit a detailed and ecologically meaningful characterization of the global ocean. As presented in Fig.~\ref{fig:2}, our eco-provinces provide a comprehensive characterization of the open ocean, with very high certainty in larger and core provinces such as the upwelling eco-province 0, the oligotrophic eco-province 3, and the provinces that encompass the Southern Ocean such as eco-provinces 5 and 6. Our eco-provinces offer a comprehensive delineation of global ocean plankton ecology and reveal detailed complexities in pelagic regions. This comprehensive representation of pelagic regions allows them to complement past characterizations such as \citet{Kavanaugh2014} seascapes and the \citet{ElHourany2024} Ocean Biomes which reveal more detail in coastal rather than pelagic regions. Furthermore, since our eco-provinces are based on phytoplankton functional type estimates on a 1/2 by 1/2 degree global ocean grid, their visible structure reveals the nuanced behavior representing the fluid boundaries and ever-changing nature of the global ocean. Our curved and intricate eco-provinces are very similar to those presented by \citet{Sonnewald2020}, but on a higher spatial resolution from 1 by 1 degree, and including a seasonal rather than annual average. Similarly, our eco-provinces are higher in resolution and complexity compared to the eight 1 by 1 degree biomes presented in \citet{HofmannElizondo2021} and the six phytoplankton communities in \citet{Kaneko2023}. Since our eco-provinces reveal a high level of ecologically informed detail in both coastal and pelagic regions while also encompassing seasonal variability, they provide potentially useful information for informing fisheries management and carbon sequestration efforts.

\subsection{Eco-Province Identification Methodology}
Our identification and prediction methodology is distinct from previous approaches \citep{Sonnewald2020, HofmannElizondo2021, Kaneko2023, Kavanaugh2014, Longhurst1995}. The current work, \citet{Sonnewald2020}, and \citet{HofmannElizondo2021} prioritize the ecological information gained from phytoplankton inputs, compared to \citet{Kavanaugh2014}, who used satellite sea surface temperature, ocean color, and chlorophyll a for an approach that is directly compatible with satellite data. \citet{Longhurst1995} also utilized satellite chlorophyll data, combined with decades of observational experience, in order to define global ocean provinces. We used the ML NEMI workflow to identify eco-provinces, which reduces the dimensionality of the seasonal phytoplankton functional type inputs using UMAP \citet{McInnes2018}. UMAP is preferred to the probabilistic approach t-SNE used in \citet{Sonnewald2020}, since the categorical cross entropy used to create the embedding in the low dimensional representation preserves both local and global, rather than just local, structures in the data. In addition, UMAP has been shown to have better computational performance compared to t-SNE \citep{McInnes2018, Jenniges2025}. \citet{Kavanaugh2014, HofmannElizondo2021} utilized a SOM for dimension reduction, which is a two layer neural network that uses competitive learning to map the input to a discrete grid \citep{Kohonen2013}. While SOMs are useful for mapping onto a fixed grid while prioritizing local topology, UMAP is preferred for capturing global structures with more efficiency when working with large datasets.

For clustering, the NEMI workflow, in addition to \citet{Kavanaugh2014, HofmannElizondo2021}, uses agglomerative clustering where the user sets the desired number of clusters. The use of agglomerative clustering is desired compared to density-based clustering in \citet{Sonnewald2020}, which does not allow the user to set the number of clusters. As a result, the approach presented in \citet{Sonnewald2020} requires additional aggregation using the Bray-Curtis dissimilarity metric \citep{Bray1957}. In addition, a comprehensive analysis of clustering algorithms has demonstrated that hierarchical clustering can better grapple with outlier sensitivity compared to DBSCAN \citep{Wani2024}.

\citet{ElHourany2024,Kaneko2023} used machine learning algorithms involving in situ genetic data to identify ocean biomes, alongside satellite data, for global ocean biome characterization. Our work is similar in that it includes a phytoplankton component, but from modeled PFT data, which allows us to characterize the global ocean, including pelagic regions, with higher detail and complexity. 

After dimension reduction and clustering in our eco-province identification methodology, the NEMI workflow increases robustness of province identification by quantifying the uncertainty using entropy. Entropy quantifies the uncertainty of each cluster assignment for each grid cell through ensembling, which allows us to choose the final eco-provinces based on the ensemble with the parameter combinations offering the least uncertainty. \citet{Sonnewald2020} also addressed stochasticity using ensembling, but our use of entropy allows a more nuanced interpretation, as it allows for us to determine the certainty with which eco-provinces are identified. Furthermore, our approach identifies the PFT composition within each eco-province (Fig.~\ref{fig:2}g) which is an advancement compared to previous characterizations that do not consider PFTs. Overall, our eco-province identification approach utilizes advanced ML methodology and detailed ecological input in order to create a unique global ocean characterization, complementing past work.

\subsection{Eco-Province Inference}
After the eco-province identification step, we predict the eco-provinces based on input data that can be remotely sensed. \citet{Mohammed2023}’s comprehensive review on ensemble deep learning demonstrates the advantage of ensemble and deep learning compared to traditional algorithms \citep{Mohammed2023}. In addition, \citet{Wyatt2022} discussed using ensembles to improve robustness of deep learning for image classification in marine environments, finding ensembles optimal in terms of robust uncertainty quantification and concluding that ensembles should be the standard for using deep learning for benthic image automation \citep{Wyatt2022}. To add another level of interpretability to our results, we utilize the XAI technique SHAP, presented in \citet{Lundberg2017}, in order to understand what inputs were important for correct and incorrect predictions of eco-provinces of interest. Using XAI for domain specific validation has been demonstrated to be insightful in \citet{Clare2022}. Overall, our inference methodology combining ensembled predictions, uncertainty quantification, and explainability, poses as a useful framework to build off of for eventual predictions in the real ocean. 

Our approach of predicting PFT-based eco-provinces based on remotely sensed input data balances the trade off between satellite compatibility and identifying the complex ecological structures present throughout the open ocean. We take inspiration from \citet{Kavanaugh2014, Kaneko2023, ElHourany2024}'s compatibility with satellite data, but add more complexity, in terms of an increased number of different eco-provinces, to the characterization of the open ocean at all latitudes. 

\subsection{Inference Method Fidelity Comparison}
 Remotely sensed phytoplankton carbon biomass has more uncertainty than chlorophyll biomass, which has more uncertainty than blue and green reflectance \citep{IOCCG2019}. The uncertainty in chlorophyll and phytoplankton carbon biomass are a result of parameter estimates in their calculation; chlorophyll being calculated using a blue green light reflectance ratio and phytoplankton carbon biomass being derived from an empirical relationship to the particle backscattering properties of water \citep{Vandermeulen2025PACE}. As we added inputs across the three DNE versions, overall accuracy increased. Most notable is the improved prediction of the North Atlantic and Arctic. In particular, going from V2 to V3 with the addition of phytoplankton carbon biomass, the prediction of the Arctic greatly improves. However, some regional locations decreased in accuracy with the addition of inputs. For example, across the three versions the prediction of the JJA Western Pacific and JJA Southern Ocean deteriorates. These examples are discussed below.

A key novelty of our work is the explainability component, which we see as an important step for use in real-world applications. One such potential application is fisheries management, where previous work has demonstrated the benefit of global ocean characterizations. For example, \citet{Palomares2020} utilized \citet{Spalding2007}'s Marine Ecoregions to categorize fishery biomass trends of exploited fish \citep{Palomares2020}. In addition, \citet{Platt2003, Koeller2009, JeanBaptisteKassi2018, Menon2019} have demonstrated that satellite-derived phytoplankton phenology can help elucidate commercial fish survival rates. Our eco-provinces offer a visually and ecologically detailed characterization of the global ocean, with uncertainty measures associated with identification and prediction. 

To demonstrate the value of interpretability in our work, we consider eco-provinces 0 and 5 which encompass coastal upwelling regions, the Eastern Tropical Pacific, and the North Atlantic. These regions are potentially useful to fisheries management stakeholders, due to their ecological composition containing high diatoms, and their location in coastal upwelling regions, the Eastern Tropical Pacific, and the North Atlantic. In addition, the connection between upwelling regions and fish has been demonstrated in \citep{Reese2011, Anguita2020}. Eco-province 5 has the largest proportion of diatoms, which are important for food web and energy transport \citep{BBres2022, Tréguer2017b}. Figs.~\ref{fig:7} and \ref{fig:8} compare the SHAP values across all three versions and inputs, where blue and green reflectance decrease in importance with the addition of chlorophyll and phytoplankton carbon biomass. Overall, chlorophyll is most influential for the prediction of these eco-provinces. The prediction accuracies of eco-provinces 0 and 5 increase going from V2 to V3 with the addition of phytoplankton carbon biomass. Since both versions contain chlorophyll, V2 could be used to prioritize lower input uncertainty while V3 could be used to prioritize prediction accuracy. 

As an additional example to highlight the explainability component of our work, we consider high coccolithophore eco-provinces 0 and 7. Coccolithophores have calcium carbonate shells, which make them important for carbon sequestration through the production and sinking of their calcium carbonate shells \citep{Li2024}. Better understanding of carbon sequestration is pertinent, due to its role in mitigating climate change \citep{USGS2008CarbonSequestration}. Figs.~\ref{fig:7} and \ref{fig:8} can be analyzed from the lens of identifying eco-provinces with high carbon sequestration potential, since there are the highest concentrations of coccolithophores in eco-provinces 0 and 7. Potential future management efforts could focus on protecting such eco-provinces in order to support carbon sequestration, while choosing the DNE that prioritizes important features while taking into account input uncertainty. Similarly to the previous example, versions 2 and 3 would both be useful as they include chlorophyll, where V2 has less input uncertainty and V3 has more output accuracy.

Figs.~\ref{fig:9} and \ref{fig:10} display the importance of medium input values to the incorrect prediction of the JJA eco-province 6 in the Southern Ocean. Blue and green reflectance are influential for this prediction, decreasing in importance across versions with chlorophyll being the most important in V2 and phytoplankton carbon biomass being the most important in V3. This incorrect prediction may relate to the input values being more mid range and close in value to those in the Arctic which prevents the DNE from distinguishing the regions. This incorrect prediction corresponds to high prediction entropy, which is expected for an incorrect prediction. For applications geared toward making predictions in the JJA Southern Ocean, V1 of the DNE would be most useful.

A cautionary tale is in the incorrect prediction of eco-province 1 as eco-province 3 in the JJA Western Pacific. Figs.~\ref{fig:9} and \ref{fig:10} reveal the role of all inputs in this incorrect prediction, most notably chlorophyll. It is important to note that in a real-world situation, there is more uncertainty in satellite-derived chlorophyll and phytoplankton carbon biomass since they are estimated based on ocean color. These increased uncertainties may complicate real-world results. In the case of our prediction using modeled data, the incorrect prediction may relate to the fact that the NEMI algorithm labeled that region as eco-province 3 in DJF but as eco-province 1 in JJA. The low uncertainty in the prediction is a reminder to keep some skepticism when using mathematical tools such as artificial intelligence, and emphasizes the importance of transparency methods such as XAI in order to identify any potential issues. V1 is best suited for applications in the JJA western Pacific.

\subsection{Shortcomings of numerical models of phytoplankton}
Since we used modeled PFT concentrations in order to identify the eco-provinces, our work serves as a conceptual framework that can be extended to be applicable in the real ocean. Although the Darwin Model is validated on in situ data, there are still substantial assumptions made throughout the model equations and uncertainties in the parameters, as would be the case when using any numerical model. The use of modeled data provides high spatial and temporal resolution of how PFTs vary throughout the global ocean, which would be impossible to obtain from sparse and disjoint in situ datasets. Furthermore, the model does not perform as well near the coasts as it does not account for finer resolution coastal dynamics. The model only captures a subset of the full range of physical and chemical forcings of the in situ ocean and only a portion of the diversity of phytoplankton types. 

Despite the caveats, our approach is necessary and valuable as a proof of concept. Future work could begin to assess what amount of in situ data would be needed for such classification in the real ocean, and could address the level of uncertainty from newer satellite algorithms for estimating phytoplankton types from space from hyperspectral sensors (e.g. NASA PACE mission) needed to use such inputs. Furthermore, our ability to infer these PFT-based eco-provinces based on remotely sensed data is especially substantial in the effort to capture ecological information from satellite data alone. We have demonstrated success predicting eco-provinces based on blue and green reflectance, and will have greater potential when using higher spectral resolution. Additional optical data will be especially useful for inferring border eco-provinces, and as we increase the number of eco-provinces identified by NEMI.

\subsection{Conclusion and future work}
Here, we have established a proof of concept for identifying global eco-provinces using NEMI, then predicting them based on modeled ocean color input data. Here we have specified 10 provinces, but the number can be changed according to the desired application. XAI reveals the importance of each modeled ocean color input for the correct and incorrect predictions of each eco-province. In this paper, we identify the importance of blue reflectance, green reflectance, chlorophyll, and phytoplankton carbon biomass for predicting eco-provinces. In addition, we delve into possible explanations for an incorrect prediction with low, compared to high, uncertainty. Future work will focus on increasing temporal resolution, in order to identify and predict eco-provinces on finer time and space scales. The temporal resolution will be increased to monthly averages in order to be more applicable to management efforts that change frequently according to the season. For increasing spatial resolution, we will consider the updated Darwin Biogeochemistry Model of Ocean color, which includes 13 ocean color wavebands at a 1/6th degree resolution rather than just blue and green reflectance at a 1/2 degree resolution. These additional inputs at a higher resolution will offer improved predictive capability. A possible expanded DNE framework could include inputs from PACE satellite data which encompasses ocean color hyperspectrally, along with chlorophyll, phytoplankton carbon biomass products, and potentially estimates of phytoplankton functional types. Our study provides a framework for eco-province identification and prediction that will be useful in developing such products for real ocean management.

\section*{Abbreviations}
DJF: December, January, February (boreal winter); DNE: dense network ensemble; JJA: June, July, August (boreal summer); ML: machine learning; NEMI: Native Emergent Manifold Interrogation; PACE: Plankton, Aerosol, Cloud and ocean Ecosystem; PFT: phytoplankton functional type; SHAP: SHapley Additive exPlanations; SOM: self-organizing map; SVM: support vector machine; UMAP: Uniform Manifold Approximation and Projection; XAI: explainable artificial intelligence.

\section*{Acknowledgments}
We acknowledge Professor James Sanchirico from the UC Davis Department of Environmental Science and Policy for helpful discussions on applications to fisheries management and Oliver Jahn, the lead computational scientist on MIT's Darwin Project, for maintaining the datasets used in our work.

\section*{Funding Statement}
This work was supported by startup funds from the University of California Davis and the UC Davis Global Affairs Grants for Advancing the United Nations Sustainable Development Goals (SDGs). The funders did not contribute to study design, data collection, analysis, publishing decision, or manuscript preparation.

\section*{Competing Interests}
The authors declare there are no conflicts of interest for this manuscript.

\section*{Data Availability Statement}
All input data are publicly available through the
\href{https://simonscmap.com/}{Simons Collaborative Marine Atlas Project (Simons CMAP)},
including the Darwin Biogeochemistry Models of
\href{https://simonscmap.dev/catalog/datasets/Darwin_Ocean_Color}{Ocean Color},
\href{https://simonscmap.dev/catalog/datasets/Darwin_Phytoplankton}{Phytoplankton},
and \href{https://simonscmap.dev/catalog/datasets/Darwin_Ecosystem}{Bulk Ecosystem Characteristics}.
All code, eco-province and entropy data, DNEs, and SHAP results are publicly available on
\href{https://github.com/CompClimate/eco-province}{the Computational Climate and Ocean Group GitHub}.

\section*{Author Contributions}
Conceptualization: M.S.; Methodology: M.S.; Software: M.M.; Formal analysis: M.M.; Investigation: M.M.; Resources: S.D.; Visualization: M.M.; Supervision: M.S.; Funding acquisition: M.S.; Writing (original draft): M.M.; Writing (review and editing): M.M., M.S., S.D. All authors approved the final submitted draft.

\section*{Supplementary Material}
Supplementary Materials and Methods, Figures S1 to S5, and Table S1 accompany this article and are included following the references below.

\bibliographystyle{plainnat}
\bibliography{EDS_version}

\clearpage
\renewcommand{\thefigure}{S\arabic{figure}}
\renewcommand{\thetable}{S\arabic{table}}
\setcounter{figure}{0}
\setcounter{table}{0}

\section{Supplementary Materials and Methods}

\subsection*{NEMI Input}
To calculate summer (June, July, August: JJA) and winter (December, January, February: DJF) averaged datasets, we calculated monthly averages for each year based on the initial three-day averages across the entire time period. After obtaining the monthly averages for each year, we found the monthly averages across the entire dataset by averaging each month across the 23 years. Finally, we found the seasonal averages by averaging the June, July, August total averages to create boreal summer datasets, and averaged December, January, February averages to find boreal winter datasets.

\subsection*{Eco-Province Identification}
NEMI uses UMAP \citep{McInnes2018} for dimension reduction, a topological data reduction method that uses categorical cross entropy for optimization, which balances representation of both the high and low dimensional structures. To determine which clusters should be combined, NEMI uses Ward's linkage which minimizes the sum of squared differences within all of the clusters. \citet{Jenniges2025} demonstrated how UMAP strengthens data associations, improving clustering outcomes. For clustering, NEMI uses hierarchical agglomeration, where each observation starts as its own cluster, then clusters are successively merged together to build nested clusters \citep{Sonnewald2023}. We repeated the embedding, clustering, entropy, and functional type composition results for 15, 20, and 25 clusters. Similar patterns were observed with more complexity. A 20 cluster example is provided here in Fig.~S1, for demonstration.

Using NEMI and assessing different numbers of eco-provinces, we were able to ensure that each cluster was coherent and that as we increased the number of clusters, the clusters became more complex as progressions of each other. This pattern of higher complexity clusters being progressions of lower complexity clusters was further validated by plotting in geographical space and relating back to PFT input concentrations. The locations and shapes of each eco-province, and the way in which the complexity progresses as we increase the number of clusters, match expectations based on PFT input concentrations. For example, we observe distinct eco-provinces for low nutrient oligotrophic regions, high nutrient coastal upwelling regions, and high-diatom regions in the JJA North Pacific and DJF Southern Ocean.

\begin{figure}[H]
\centering
\includegraphics[width=0.95\linewidth]{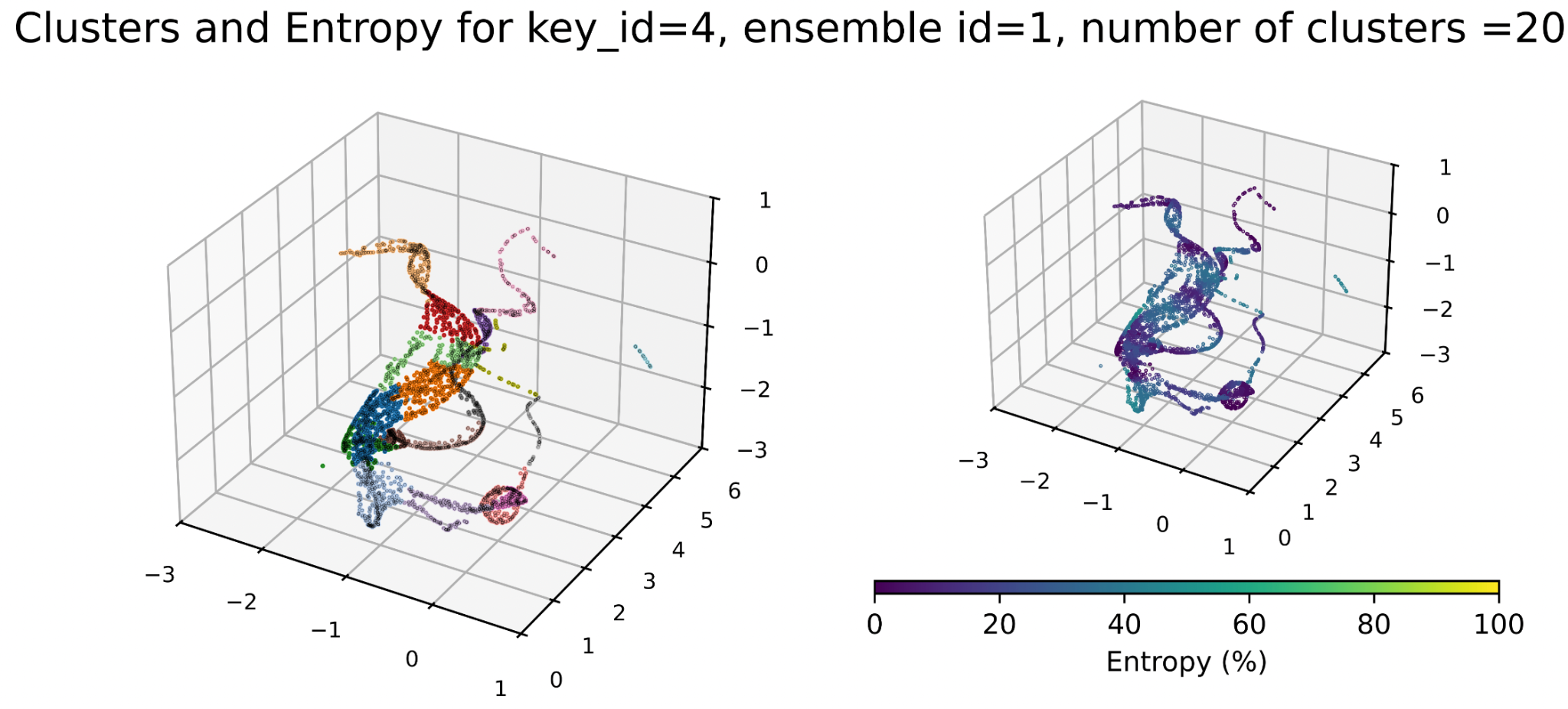}\\[8pt]
\includegraphics[width=0.8\linewidth]{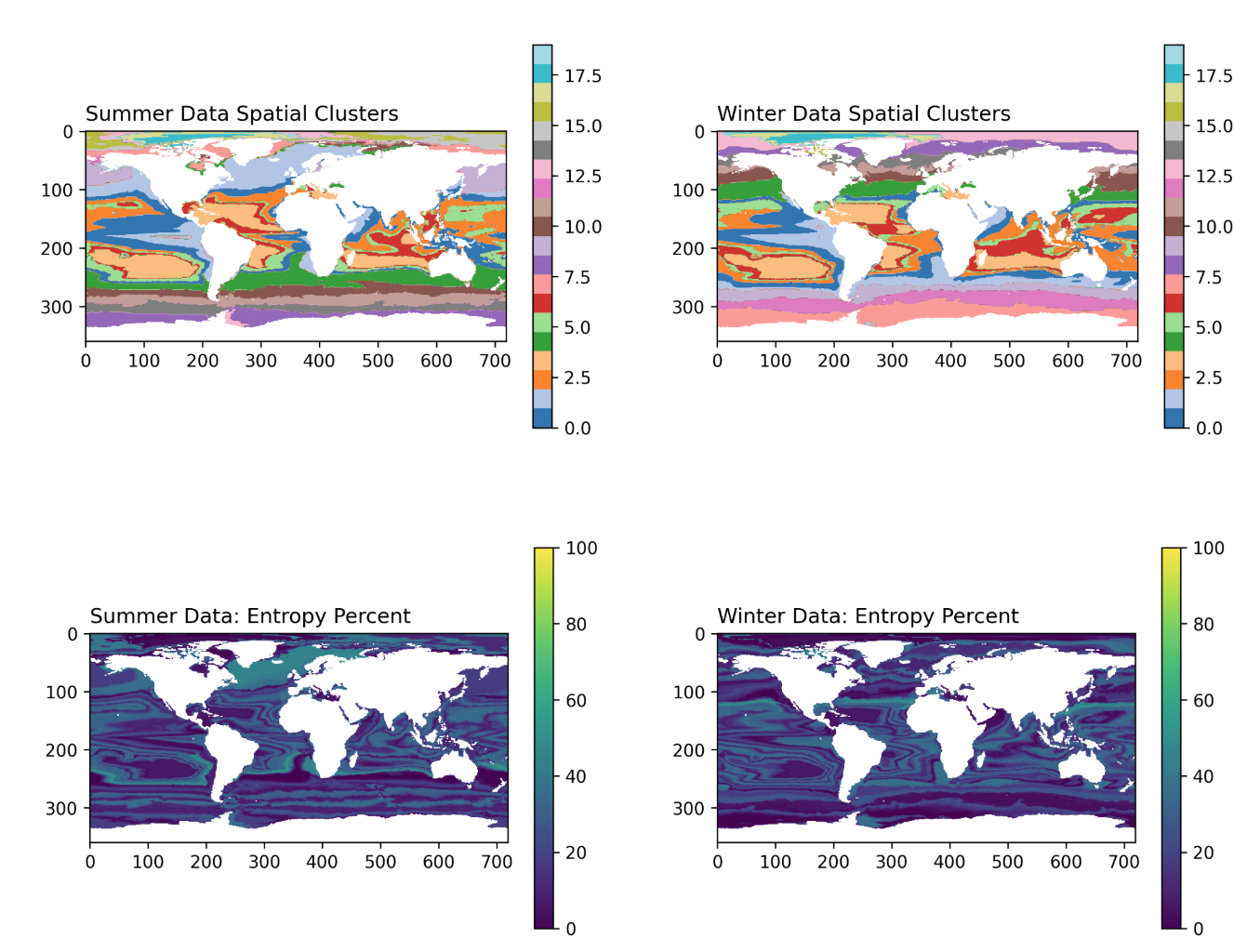}
\caption{\textbf{20-Cluster Example.} Example of an eco-province characterization with 20 eco-provinces (clusters) and the corresponding entropy percentage.}
\label{fig:S1}
\end{figure}

Fig.~S2 demonstrates the neural network train, validation, and test split with testing on the western Atlantic and validation on the eastern Atlantic.

\begin{figure}[H]
\centering
\includegraphics[width=0.95\linewidth]{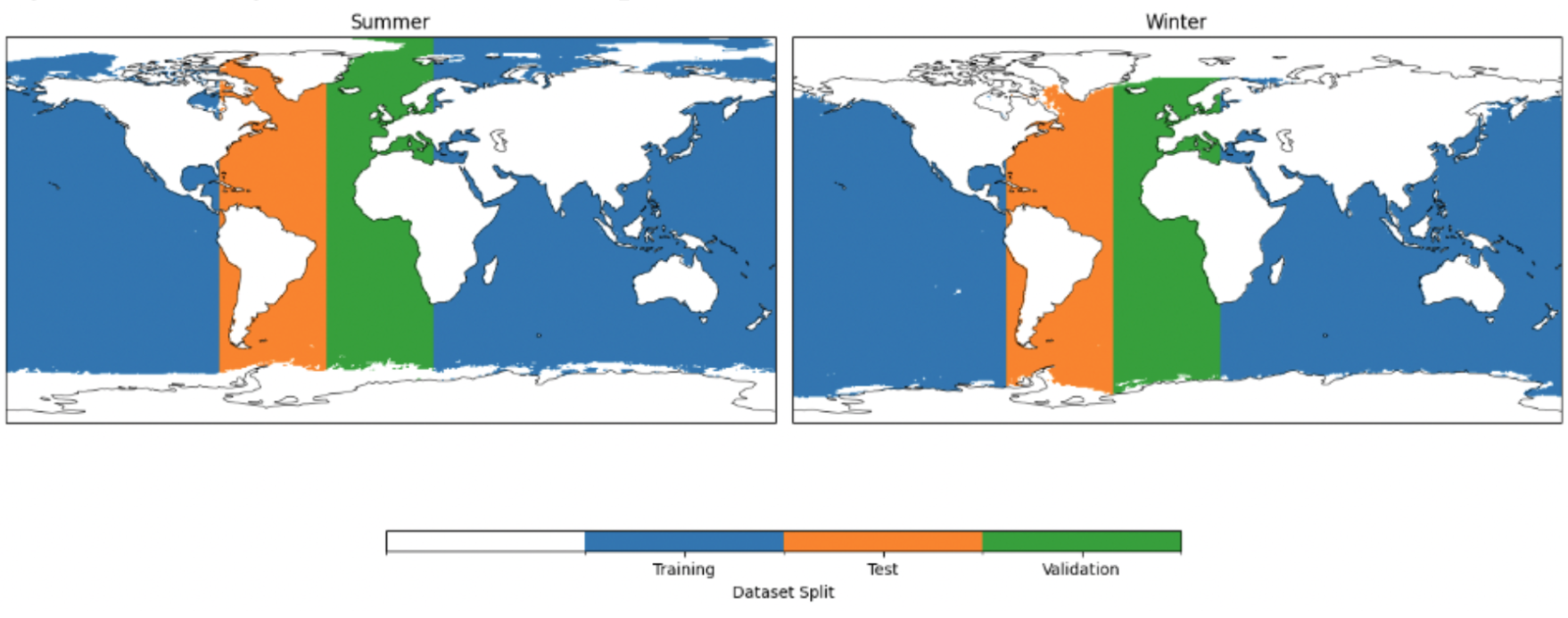}
\caption{\textbf{Training, validation, and test split.} Fig.~S2 demonstrates the vertical split used to train, validate, and test the neural network with input data averaging across 1993-2004. For an extra level of validation, we then tested the network on the entire globe for 2005.}
\label{fig:S2}
\end{figure}

In each of the versions, the inputs are passed into four hidden layers, each with a hyperbolic tangent activation function. The output layer uses a softmax activation function to predict eight resulting eco-provinces. The neural network uses categorical cross entropy for the loss function and trains for 100 epochs.

Table~S1 compares the accuracy of each version for boreal summer (June, July, August: JJA) and boreal winter (December, January, February: DJF), in addition to the average accuracy.

\begin{figure}[H]
\centering
\includegraphics[width=0.4\linewidth]{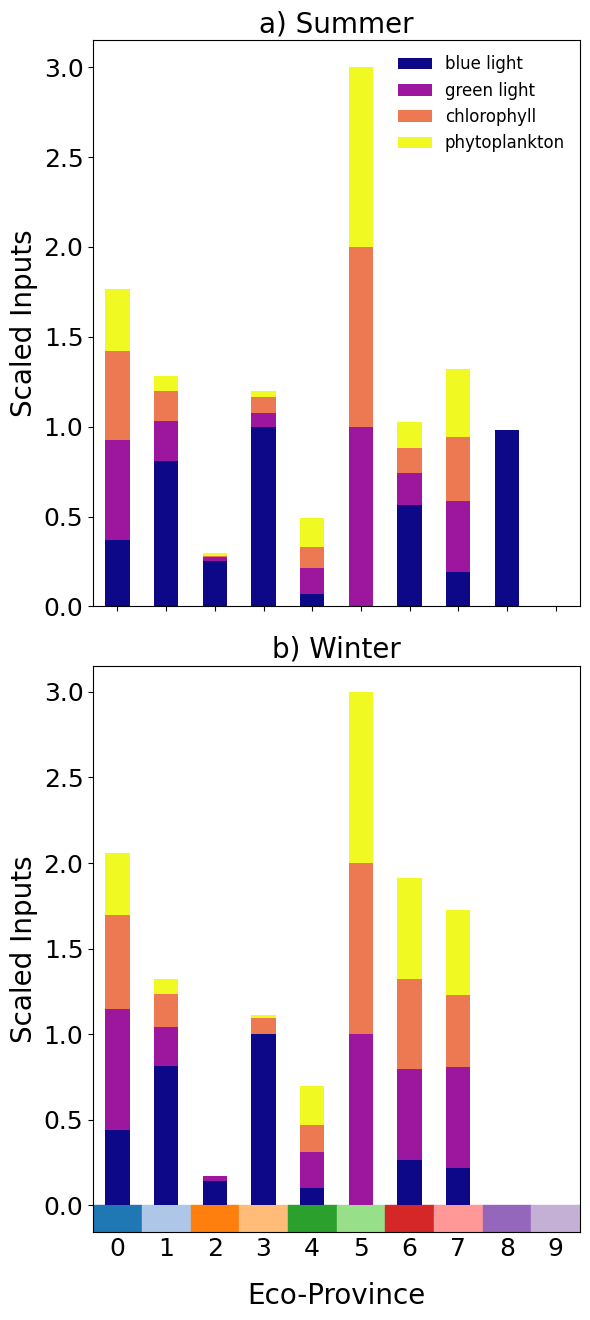}
\caption{\textbf{DNE Input Composition}. Fig.~S3 presents the scaled composition of blue reflectance, green reflectance, chlorophyll, and phytoplankton carbon biomass as inputs to the V3 DNE, broken up according to each eco-province. This composition allows us to identify the input composition for each different spatial region, in order to understand which inputs contributed to specific results. In general, either high or low input concentrations are more informative for DNE predictions compared to moderate concentrations.}
\label{fig:S3}
\end{figure}

\begin{table}[H]
\centering
\includegraphics[width=0.8\linewidth]{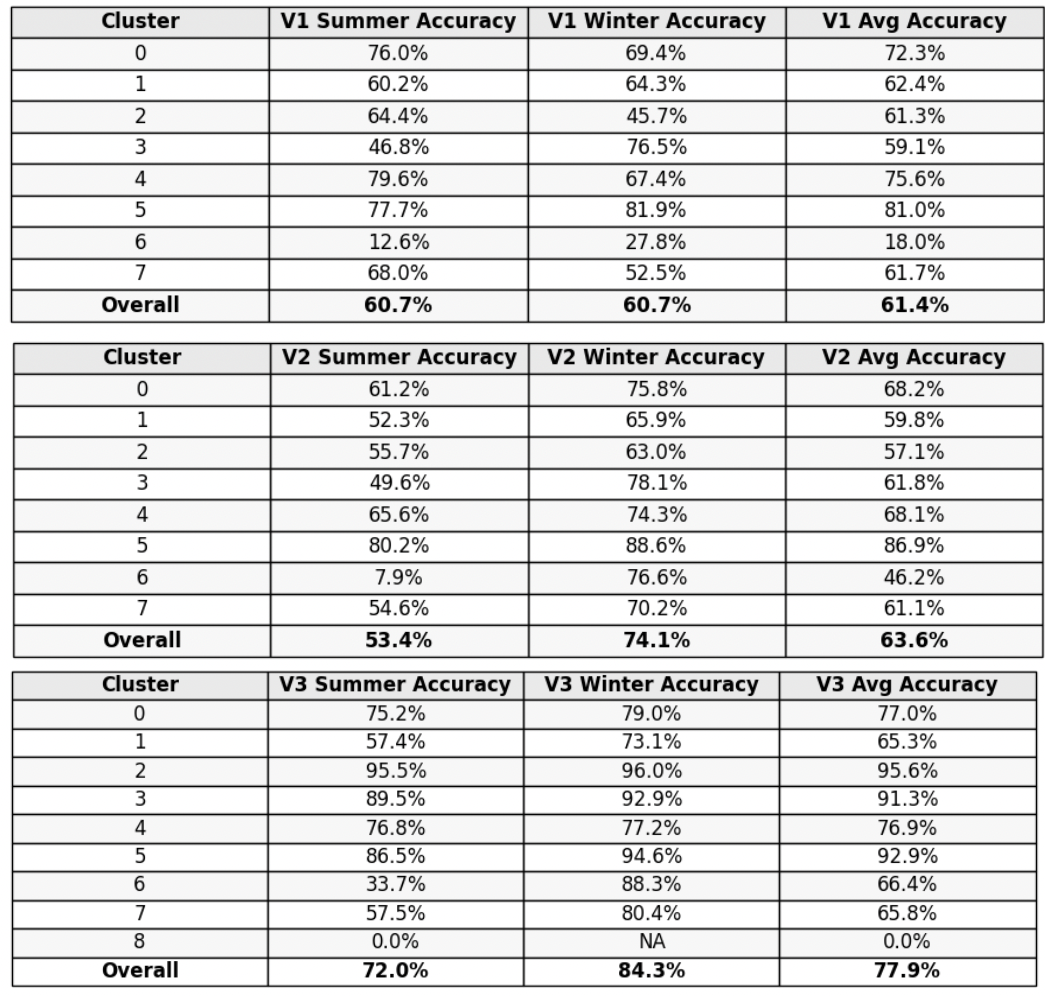}
\caption{\textbf{Summer (JJA), Winter (DJF), and Average Accuracy Across Versions.} Table~S1 demonstrates the variety in accuracy among the three versions, revealing the increase in overall performance with the addition of inputs with some nuance in performance when considering each cluster individually.}
\label{tab:S1}
\end{table}

Fig.~S4 compares the ensemble losses and accuracies across the three versions.

\begin{figure}[H]
\centering
\includegraphics[width=0.95\linewidth]{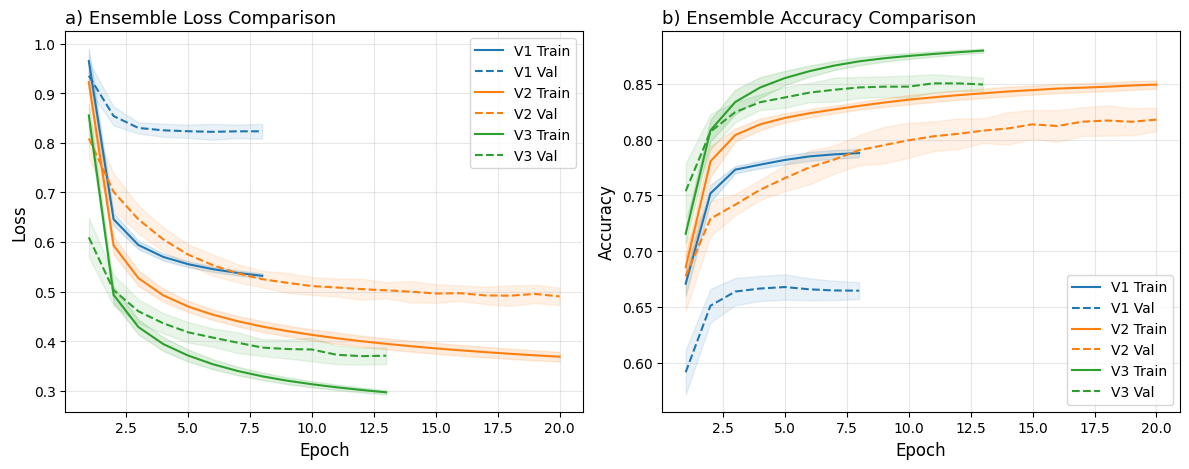}
\caption{\textbf{Ensemble loss and accuracy comparison across three versions.}}
\label{fig:S4}
\end{figure}

\begin{figure}[H]
\centering
\includegraphics[width=0.95\linewidth]{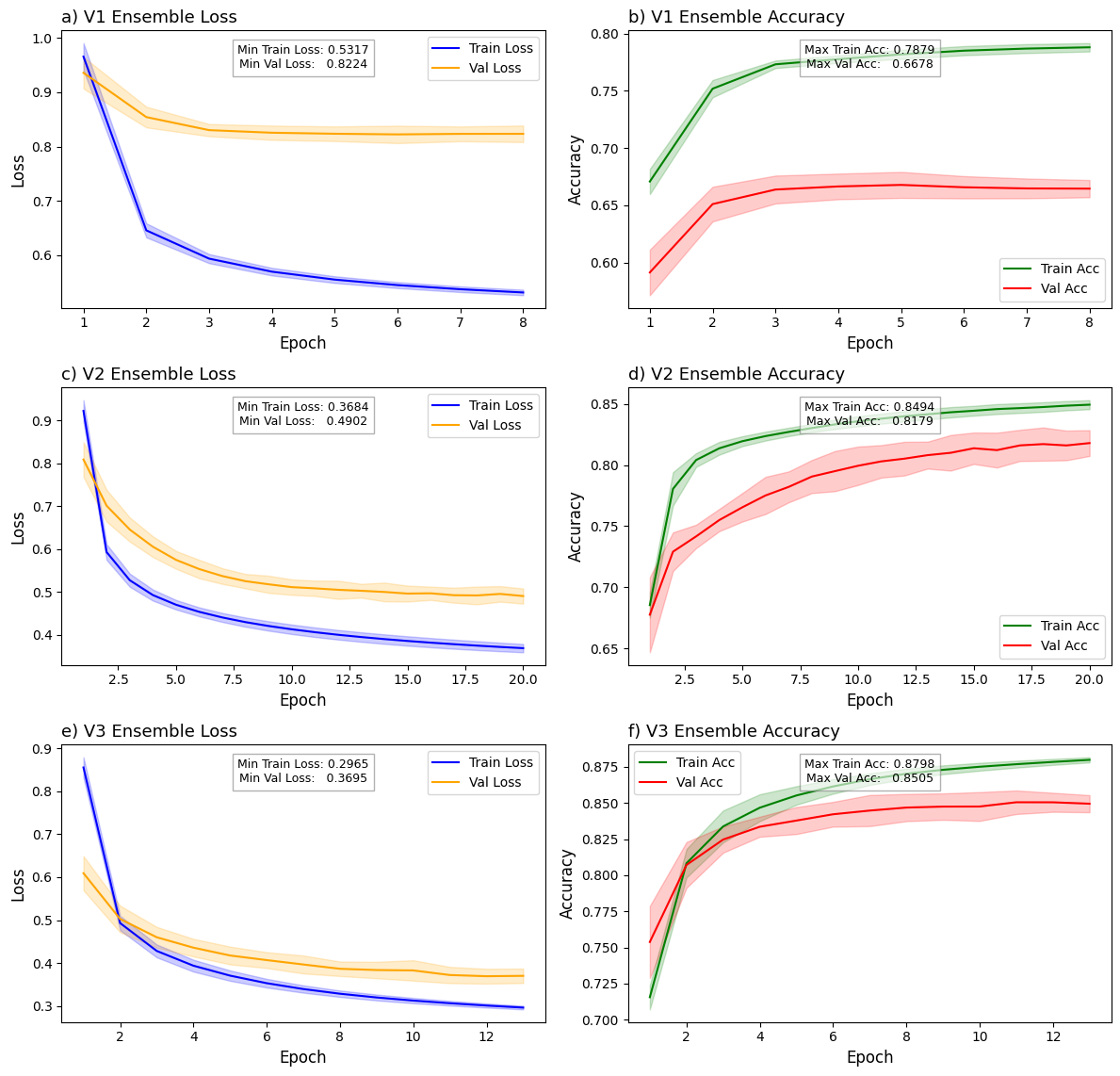}
\caption{\textbf{Ensemble loss and accuracy comparison across three versions, with minimum loss and maximum accuracy values.} Fig.~S5 also compares the ensemble losses and accuracies across the three versions, with minimum loss values and maximum accuracies.}
\label{fig:S5}
\end{figure}

\end{document}